\documentclass[useAMS,usenatbib,usegraphicx]{mn2e}

\usepackage{times}

\title[Dust emission from forming galaxies]{
  A model for the infrared dust emission from forming galaxies}

\author[T.\ T. Takeuchi et al.]
  {Tsutomu T.~Takeuchi,$^{1}$\thanks{E-mail: tsutomu.takeuchi@oamp.fr.
    }\thanks{
    Postdoctoral Fellow of the Japan Society for the Promotion of Science for 
    research abroad.}
   Takako T. Ishii,$^{2}$\thanks{
    Postdoctoral Fellow of the Japan Society for the Promotion of Science.}
\newauthor
   Takaya Nozawa,$^3$
   Takashi Kozasa,$^3$
   and
   Hiroyuki Hirashita$^{4,5,6}\ddagger$
    \\
$^1$Laboratoire d'Astrophysique de Marseille, Traverse du Siphon BP8,
       13376 Marseille Cedex 12, France\\
$^2$Kwasan Observatory, Kyoto University, Yamashina-ku, Kyoto
       607--8471, Japan\\
$^3$Division of Earth and Planetary Sciences, Hokkaido University, Sapporo
       060--0810, Japan\\
$^4$Division of Particle and Astrophysical Sciences, Nagoya University, 
       Nagoya 464--8602, Japan\\
$^5$SISSA-ISAS, International School for Advanced Studies, Via Beirut 4, 
       34014 Trieste, Italy\\
$^6$Center for Computational Sciences, University of Tsukuba,
       Tsukuba, Ibaraki 305--8577, Japan
}

\date{Released 2002 Xxxxx XX}

\def\sbs{SBS~0335$-$052}
\def\izw{I~Zw~18}
\def\cb{MS~1512$-$cB58}

\begin{document}

\label{firstpage}

\maketitle

\begin{abstract}
Dust plays various important roles in galaxy formation and evolution.
In the early epoch of galaxy evolution, dust is only supplied by 
supernovae (SNe).
With the aid of a new physical model of dust production by SNe
developed by \citet{nozawa03}(N03), we constructed a model of dust 
emission from forming galaxies on the basis of the theoretical framework of 
\citet{takeuchi03a} (T03).
N03 showed that the produced dust species depends strongly on the mixing 
within SNe. 
We treated both unmixed and mixed cases and calculated the infrared (IR) 
spectral energy distribution (SED) of forming galaxies for both cases.
Our model SED is less luminous than the SED of T03 model by
a factor of 2--3.
The difference is due to our improved treatment of UV photon
absorption cross section, as well as different grain size and species
newly adopted in this work.
The SED for the unmixed case is found to have an enhanced near to mid-IR 
(N--MIR) continuum radiation in its early phase of the evolution 
($\mbox{age} \la 10^{7.25}\;$yr) compared with that for the mixed case.
The strong N--MIR continuum is due to the emission from Si grains, 
which only exist in the species of the unmixed dust production.
We also calculated the IR extinction curves for forming galaxies.
N03 dust was found to yield a smaller extinction than that of T03 model.
For the unmixed case, near-IR (NIR) extinction is dominated by large
grains of Si and amorphous carbon, 
and silicate features are less prominent compared to the curve given by T03.
On the contrary, the extinction curve of the mixed case has a similar shape
with that of T03.
Then we calculated the SED of a local starbursting dwarf galaxy \sbs.
Our present model SED naturally reproduced the strong N--MIR continuum 
and the lack of cold FIR emission of \sbs. 
We found that only the SED of unmixed case can reproduce the NIR continuum
of this galaxy.
We then made a prediction for the SED of another typical star-forming dwarf,
\izw.
The MIR continuum of \izw\ is expected to be much weaker than that of T03 SED.
We also presented the evolution of the SED of LBGs.
Finally, we discussed the possibility of observing forming galaxies at 
$z \ga 5$.
\end{abstract}

\section{Introduction}

In spite of the recent vast progress both in observational and theoretical
studies, our understanding of the physics of galaxy formation and evolution 
is still far from sufficient.

The cosmic star formation (SF) history, introduced by \citet{tinsley80} and 
developed by subsequent studies \citep[e.g.,][]{lilly96,madau96}, 
has always drawn much attention.
Now the observations reach up to $\sim 6\mbox{--}8$ 
\citep[e.g,][]{stanway03,bouwens04a,bouwens04b}.
In such studies, the role of dust has been increasingly recognized when we
try to understand the evolution of galaxies in the context of cosmic star 
formation history, because dust grains absorb stellar light and re-emit 
it in the far infrared (FIR). 
Even a small amount of dust can lead to a significant underestimation of 
the star formation rate (SFR) \citep{steidel99,adelberger00}.
Indeed, there is another extreme category of high-$z$ galaxies which have
large amount of dust and are extremely luminous in the FIR and submillimetre 
(submm) wavelengths \citep[e.g.,][]{hughes98,eales03}.
Heavily hidden SF is suggested in these galaxies 
\citep[e.g.,][]{takeuchi01a,takeuchi01b,totani02}.

Further, from physical point of view, dust grains are 
one of the fundamental ingredients in the activity of galaxies.
Since they are formed in a variety of environments ranging from 
explosive ejecta of novae and supernovae to the outflowing gas of evolved
low-mass stars, the dust formation is closely related to the SF
activity \citep{dwek98}.
Moreover, the existence of dust is crucial in the physical process of 
galaxy formation and evolution through the formation of molecular hydrogen
\citep[e.g.,][]{hirashita02b}.
The dust itself plays a leading part in the physics of SF activity
in galaxies in the whole period of the cosmic history.

Then, how about very young galaxies in the early universe at $z\ga 5$?
It is often assumed, without deliberation, that the effect of dust is 
negligible for such young galaxies, because of their low metallicities.
There is, however, a good counter-intuitive example in the local Universe: 
a local dwarf star-forming 
galaxy \sbs\ has a very young stellar age ($\la 10^{7}\,\mbox{yr}$) and 
low metallicity ($1/41\,Z_\odot$), but has a heavily embedded active star 
formation and strong continuum radiation in N--MIR wavelength regime 
\citep{dale01,hunt01}.
In addition, there are increasing number of observations suggesting the
existence of dust in high-$z$ systems such as Lyman $\alpha$ systems 
\citep[e.g.,][]{ledoux02,ledoux03} or QSOs \citep[e.g.,][]{bertoldi03}.

{}To produce dust effectively in such a young system,
the dust enrichment should have occurred primarily in the ejecta of supernovae
(SNe), especially Type II supernova (SN II) explosions\footnote{
More specifically, core-collapse supernovae.}, because the lifetime 
of the progenitor is short enough ($\sim 10^6\,\mbox{yr}$) 
\citep[e.g.,][]{dwek80,kozasa89}.
Recent theoretical studies claim the possibility of very massive stars for
the first generation stars, which end their lives as 
pair-instability supernovae (PISNe) \citep{heger02,umeda02}.
We should also take into account for the PISNe to study the very early 
evolution of dust.
Recently, the dust formation in SNe ejecta has been observationally supported
\citep[e.g.,][]{douvion01,morgan03a,dunne03}.

For investigating the properties of dust in details, theoretical 
predictions for the amount and composition of dust are required.
So far, some theoretical models of dust production by SNe have been developed.
\citet[][hereafter TF01]{todini01} showed that the dust mass produced by 
a SN II is 0.1--0.4 $M_\odot$ applying the theory of nucleation and grain 
growth by \citet{kozasa87}. 
In the calculations, they assumed an adiabatic cooling in the ejecta
with uniform gas density as well as elemental composition within He-core.
They also found that SNe form amorphous carbon with size around
300 \AA\ and silicate grains around 10--20 \AA.
\citet{schneider04} extended the progenitor mass range to the regime of PISNe
(140--260 $M_\odot$) and found that 10--60 $M_\odot$ of dust forms per PISN. 
The grain radius depends on the species and is distributed from 0.001 to
0.3 $\mu$m. 

Recently, \citet[][]{nozawa03} (hereafter N03) carefully took into account the
radial density profile and the temperature evolution in the calculation of the 
dust mass in the ejecta of SNe II and PISNe.
N03 showed that the produced dust species depends strongly on the mixing 
within SNe. 
In particular, carbon dust is not produced in the mixed case, 
because the carbon and oxygen are mixed and combined to form CO molecules. 
On the contrary, it forms in unmixed SN, since there is a carbon-rich region
at a certain location in the ejecta of SNe. 
In addition, N03 predicts a dust mass larger than that of TF01 for SNe II.

\begin{figure*}
\centering\includegraphics[angle=90,width=0.9\textwidth]{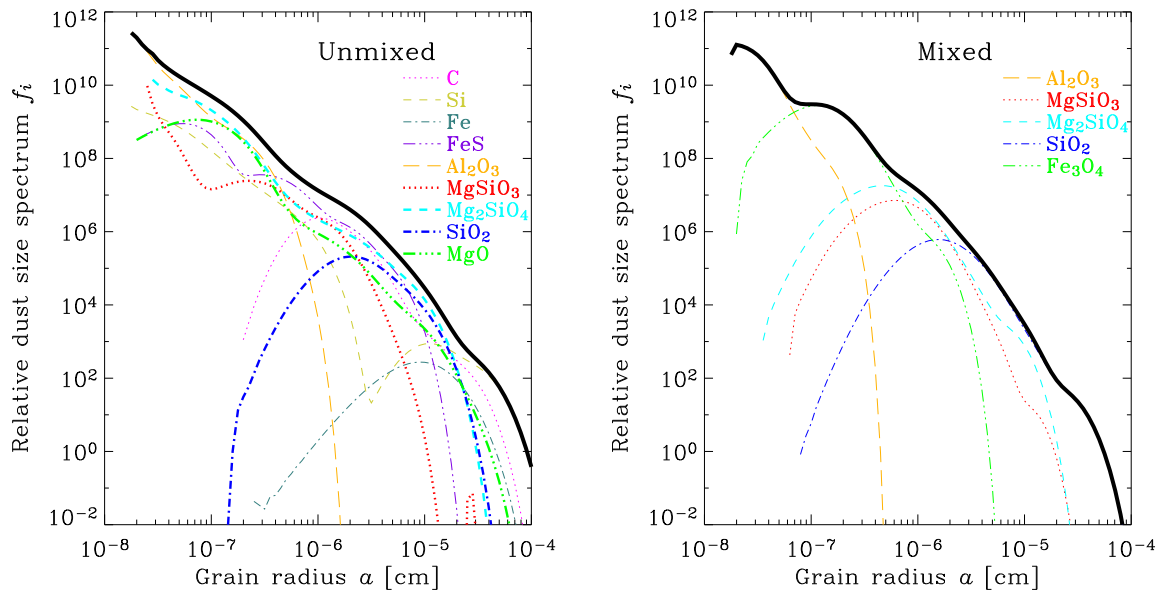}
\caption{Dust size distributions of type II supernova model with the
progenitor mass of $20\;M_\odot$ by \citet[][]{nozawa03}.
The thick lines are the total dust size distribution function.
Labels `Unmixed' and `Mixed' indicate the adopted assumption of 
mixing of elements in the ejecta.
We assumed $20\,M_\odot$ as a representative
for the supernova progenitor mass in this work.
}\label{fig:dust_size}
\end{figure*}

The amount of dust produced in SNe is thus still a matter of debate. 
\citet{morgan03b} adopted a dust production smaller than those of TF01 or N03
in their theoretical calculation, in line with IR observations at the time.
On the other hand, various observational efforts have been devoted to 
constrain the dust production efficiency of SNe 
\citep[e.g.,][]{arendt99,dunne03,morgan03a,green04,hines04,krause04,wilson05}.
These results range from $0.003\;M_\odot$ to $1.0\;M_\odot$ per SN, 
and N03 dust production seems consistent with the upper range of these
observational constraints.
However, in these observational evaluations, a large uncertainty still remains
due to the uncertainties in the estimation of dust mass in Cas~A caused by
the foreground contamination \citep{krause04}, although the evidence for dust 
in some SN remnants (SN1987A and Kepler) is still pointing towards SN origin.
Thus, further close interaction between theoretical and observational works 
is required to overcome the uncertainties and obtain a reasonable picture
of dust production in SNe.

In order to examine the dust properties of such high-$z$ galaxies, the most
direct observable is their spectral energy distribution (SED),
especially at the FIR wavelengths.
\citet{hirashita02a} modeled the evolution of FIR luminosity and 
dust temperature in a young starburst on the basis of SNe II grain 
formation model of TF01.
\citet{takeuchi03a} (T03) subsequently constructed a model of infrared
(IR) spectral energy distribution (SED) of galaxies starting from the model
of \citet{hirashita02a}.
T03, for the first time, properly consider the dust size distribution 
peculiar to the very early stage of galaxy evolution in the model of 
the IR SED of very young galaxies, and successfully reproduced 
the peculiar MIR SED of \sbs\ (though their result seems to have overestimated
the FIR continuum).
\citet{takeuchi04b} (T04) applied the T03 model to the Lyman-break galaxies.
They found that the suggested hot dust in the Lyman-break galaxies
\citep[e.g.,][]{ouchi99,chapman00,sawicki01} can be naturally explained by 
the T03 model.

Since these works are based on TF01 conjecture, the dust formation model 
used is based on the classical nucleation theory \citep{feder66}.
Recently, however, the importance of nonequilibrium (non-steady state) 
effects on the dust grain formation has been recognized in various 
astrophysical contexts \citep[e.g., ][N03]{gail84,tanaka02}.
In addition, N03 found that the radial density profile of the SN progenitor 
and the temperature evolution of the ejecta also affect the dust grain 
formation.
Hence, now is the time to take into account these concepts to further
investigate the dust emission from young galaxies.
In this work, we construct a new model of IR SED of extremely young galaxies
based on N03 SN dust formation model.
Starting from the size distribution and the amount of dust predicted by N03, 
we calculate the dust emission model by extending the T03 model to treat
multiple dust species.

The paper is organized as follows:
In \S\ref{sec:model} we explain the framework of our SED model.
We present the basic result, the evolution of the SED of extremely
young galaxies, in \S\ref{sec:results}.
Related discussions on local star-forming dwarf galaxies and high-$z$ galaxies
will be in \S\ref{sec:discussion}.
\S\ref{sec:conclusion} is devoted to our conclusions.
Throughout this paper, we use a cosmological parameter set of 
$(h,\Omega_0,\lambda_0)=(0.7,0.3,0.7)$, where 
$h\equiv H_0/100 \;[\mbox{km\,s}^{-1}\mbox{Mpc}^{-1}]$.

\section{SED Model for Forming Galaxies}\label{sec:model}

\begin{table}
\begin{center}
\begin{minipage}{0.45\textwidth}
\caption{Dust grain species.}
\begin{tabular}{@{}lcccc@{}}\hline
Species & Unmixed & Mixed &  Density ($\rho_i$) & 
Ref\footnote{References for optical constants: 
(1) \citet{edo85}, 
(2) \citet{edward85},
(3) \citet{philipp85}, 
(4) \citet{lynch91},
(5) \citet{semenov03}, 
(6) \citet{mukai89},
(7) \citet{toon76},
(8) \citet{roessler91}, and 
(9) \citet{dorschner95}.
}\\
              &           &         & [$\mbox{g}\,\mbox{cm}^{-3}$] & \\\hline
C             & $\circ$   &         & 2.28 & 1 \\
Si            & $\circ$   &         & 2.34 & 2 \\
SiO$_2$       & $\circ$   & $\circ$ & 2.66 & 3 \\
Fe            & $\circ$   &         & 7.95 & 4 \\
FeS           & $\circ$   &         & 4.87 & 5 \\
Fe$_3$O$_4$   &           & $\circ$ & 5.25 & 6 \\
Al$_2$O$_3$   & $\circ$   & $\circ$ & 4.01 & 7 \\
MgO           & $\circ$   &         & 3.59 & 8 \\
MgSiO$_3$     & $\circ$   & $\circ$ & 3.20 & 9 \\
Mg$_2$SiO$_4$ & $\circ$   & $\circ$ & 3.23 & 5 \\ \hline
\end{tabular}
\label{tab:dust_species}
\end{minipage}
\end{center}
\end{table}

\subsection{Species and size distribution of dust grains produced by SNe II 
}\label{subsec:dust_species}

\subsubsection{Dust production model of Nozawa et al.\ (2003) (N03)}

N03 investigated the formation of dust grains in the ejecta of Population 
III SNe (SNe II and PISNe, whose progenitors are initially metal-free). 
As we mentioned above, they treat some aspects which TF01 have not taken 
into account:
(i) the time evolution of gas temperature is calculated by solving 
the radiative transfer equation including the energy deposition of 
radioactive elements.
(ii) the radial density profile of various metals is properly
considered, and 
(iii) unmixed and uniformly mixed cases in the He core are considered.
In the unmixed case, the original onion-like structure of elements is 
preserved, and in the mixed case, all the elements are uniformly mixed
in the helium core.

It should be mentioned here again that TF01 assumed an adiabatic cooling
in the ejecta and adjusted the adiabatic index $\gamma$ to 1.25, referring to
the formation episode of dust grains observed in SN~1987A.
As pointed out by \citet{kozasa89}, the condensation time as well as 
the resulting average size of dust grains strongly depend on the value of 
$\gamma$.
SN~1987A is somehow a peculiar SN in the sense that the progenitor is not
a red supergiant but blue supergiant:
see \citet{arnett89} for the details.
Using SN~1987A as a template may not be appropriate in comparisons to 
supernovae as a whole.
Therefore in this paper, we use the result of N03 as a standard model.

We should note that N03 also assume the complete formation of CO and SiO 
molecules, neglecting the destruction of those molecules, i.e., 
no carbon-bearing grain condenses in the region of ${\rm C/O}<1$ and 
no Si-bearing grain, except for oxide grains, condenses in the region of 
${\rm Si/O}<1$.
The formation of CO and SiO may be incomplete because of the destruction by
energetic electron impact within SNe. 
TF01 treat both formation and destruction of CO and SiO, finding that
both are mostly destroyed. 
The decrease of CO leads to the formation of carbon grains, which could 
finally be oxidised with available oxygen. 
The destruction of SiO could decrease the formation of grains composed of 
SiO$_2$, MgSiO$_3$, and Mg$_2$SiO$_4$, and increase other oxidised grains 
and Si grains.
Observationally, it is still a matter of debate if CO and SiO are efficiently 
destroyed or not.
For the detailed discussions on this issue, see Appendix~B of N03.

\begin{figure*}
\centering\includegraphics[angle=90,width=0.9\textwidth]{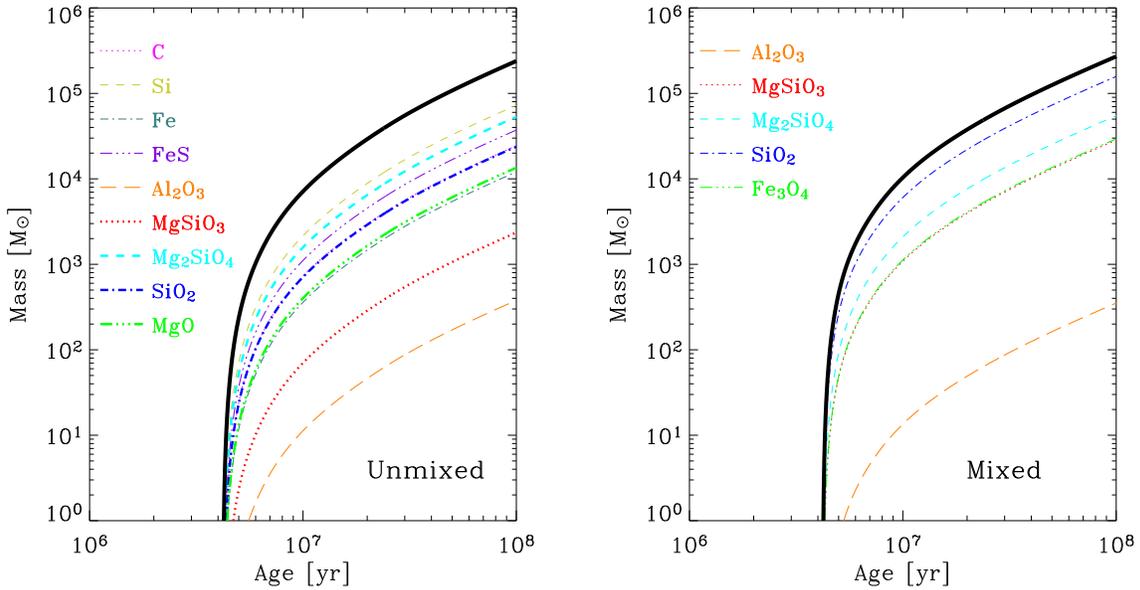}
\caption{
Evolution of the mass for each grain species.
The thick lines indicate the total amount of dust mass.
We assumed the closed box model and a constant star formation rate.
The mass fraction of each grain species is constant in time and is 
given by the size distribution function (Fig.~\ref{fig:dust_size}).
}\label{fig:dust_amount}
\end{figure*}

\subsubsection{Dust grain species produced by N03 model}

In the unmixed ejecta, a variety of grain species
(Si, Fe, Mg$_2$SiO$_4$, MgSiO$_3$, MgO,
Al$_2$O$_3$, SiO$_2$, FeS,
and C) condense, and in the mixed ejecta, in contrast, only oxide
grains (SiO$_2$, MgSiO$_3$, Mg$_2$SiO$_4$, Al$_2$O$_3$, and Fe$_3$O$_4$) form.
This is because carbon atoms are consumed to form CO molecules. 
We summarize the species formed in SNe in Table \ref{tab:dust_species},
where the species marked with a circle are relevant for unmixed
and mixed SNe.
The size of the grains spans a range of three orders of magnitude, 
depending on the grain species. 
The size spectrum summed up over all the grain species has a very 
broad distribution, and very roughly speaking, it might be approximated by
a power law.
This size distribution is different from that of the SN II calculation of 
TF01, which has a typical sizes of 300 \AA\ for amorphous carbon and
10--20 \AA\ for oxide grains. 

In this work, we adopt the representative progenitor mass of SNe II as 
$20\;M_\odot$. 
N03 have shown that the size distribution of each grain species is almost
independent of the progenitor mass, if the SN type is fixed (i.e., SN II or 
PISN). 
We examine the unmixed and mixed cases.
The size distributions of dust grains in the mixed and unmixed cases 
calculated by N03 are shown in Figure~\ref{fig:dust_size}.
We used these size distributions after binning with a bin width of 0.2~dex
for our calculations.
Throughout this work, we assume a uniform and spherical grain.
It should be mentioned that a different shape of dust grains in
SN is suggested \citep[e.g.,][]{dwek04}.

\subsection{Star formation, chemical evolution, and dust production}

\begin{table}
\begin{center}
\begin{minipage}{0.45\textwidth}
\caption{Mass fraction of dust species (unmixed).}
\begin{tabular}{@{}lc@{}}\hline
Species & Mass fraction \\ \hline
C             & 0.102   \\
Si            & 0.303   \\
Fe            & 0.050   \\
FeS           & 0.156   \\
Al$_2$O$_3$   & 0.002   \\
MgSiO$_3$     & 0.010   \\
Mg$_2$SiO$_4$ & 0.222   \\
SiO$_2$       & 0.099   \\
MgO           & 0.056   \\ \hline
\end{tabular}
\label{tab:mass_fraction_unmixed}
\end{minipage}
\end{center}
\end{table}

\begin{table}
\begin{center}
\begin{minipage}{0.45\textwidth}
\caption{Mass fraction of dust species (mixed).}
\begin{tabular}{@{}lc@{}}\hline
Species & Mass fraction \\ \hline
Al$_2$O$_3$   & 0.001 \\
MgSiO$_3$     & 0.105 \\
Mg$_2$SiO$_4$ & 0.202 \\
SiO$_2$       & 0.583 \\
Fe$_3$O$_4$   & 0.108 \\ \hline
\end{tabular}
\label{tab:mass_fraction_mixed}
\end{minipage}
\end{center}
\end{table}

\begin{figure*}
\centering\includegraphics[width=0.3\textwidth]{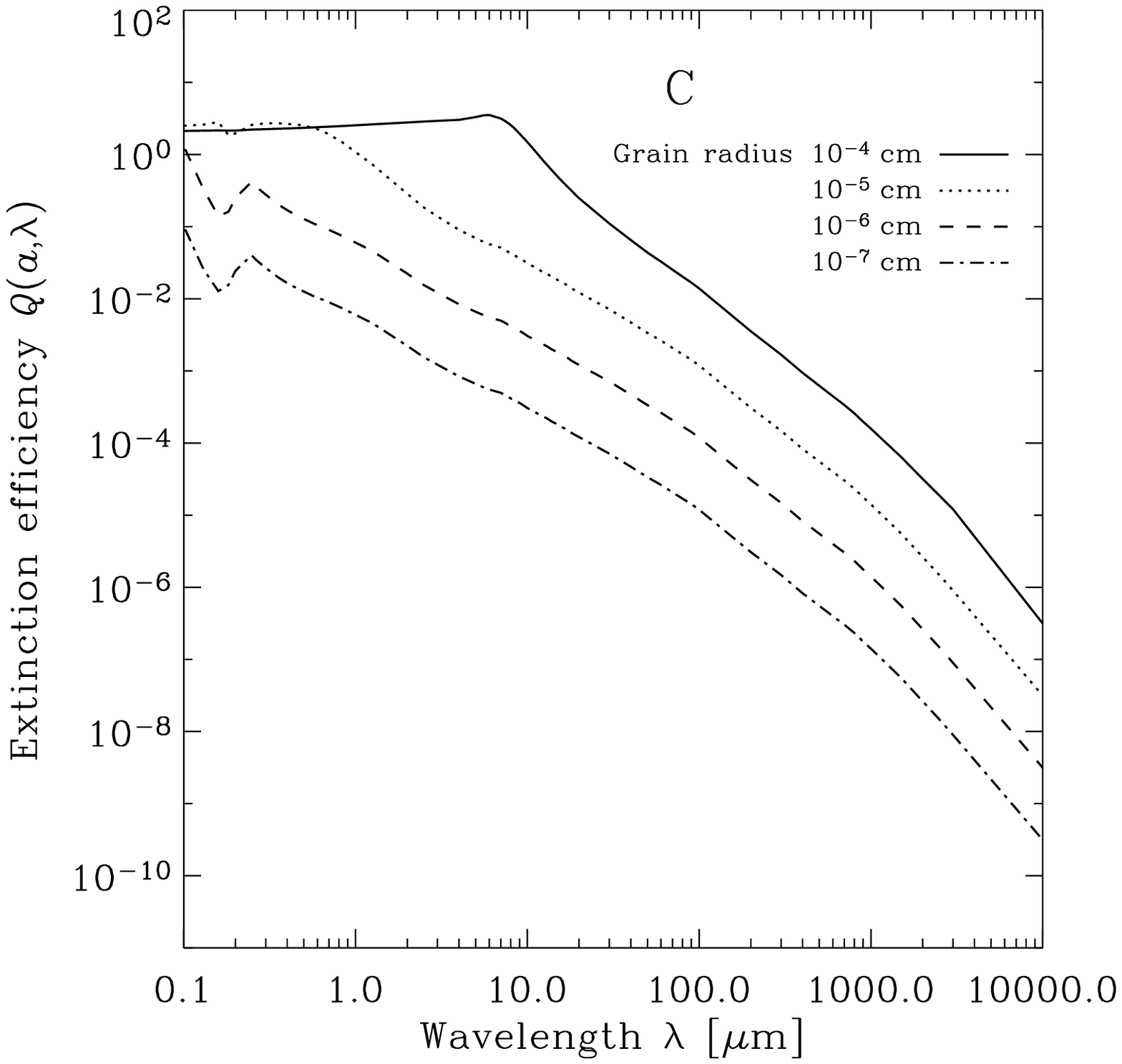}
\centering\includegraphics[width=0.3\textwidth]{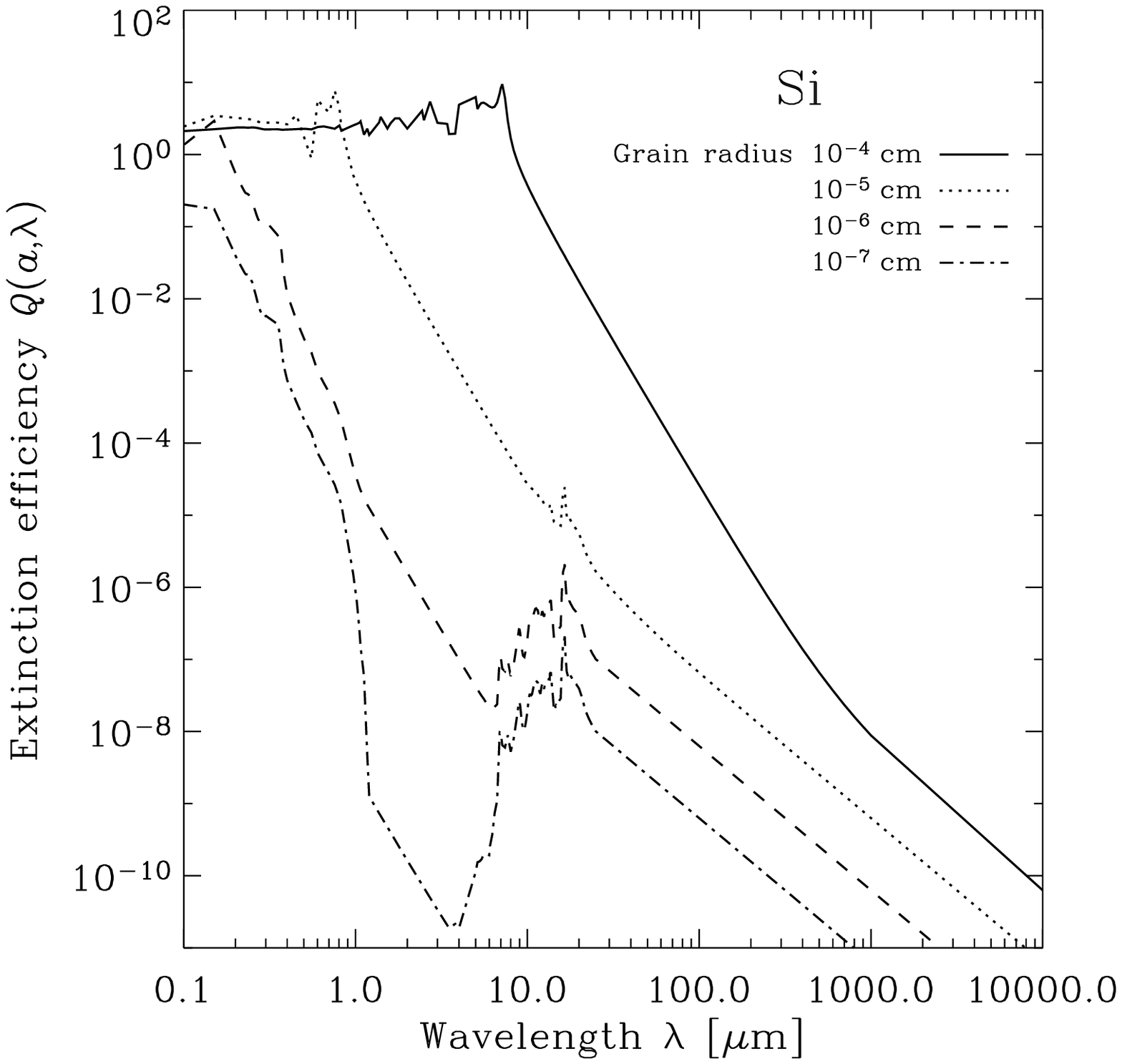}
\centering\includegraphics[width=0.3\textwidth]{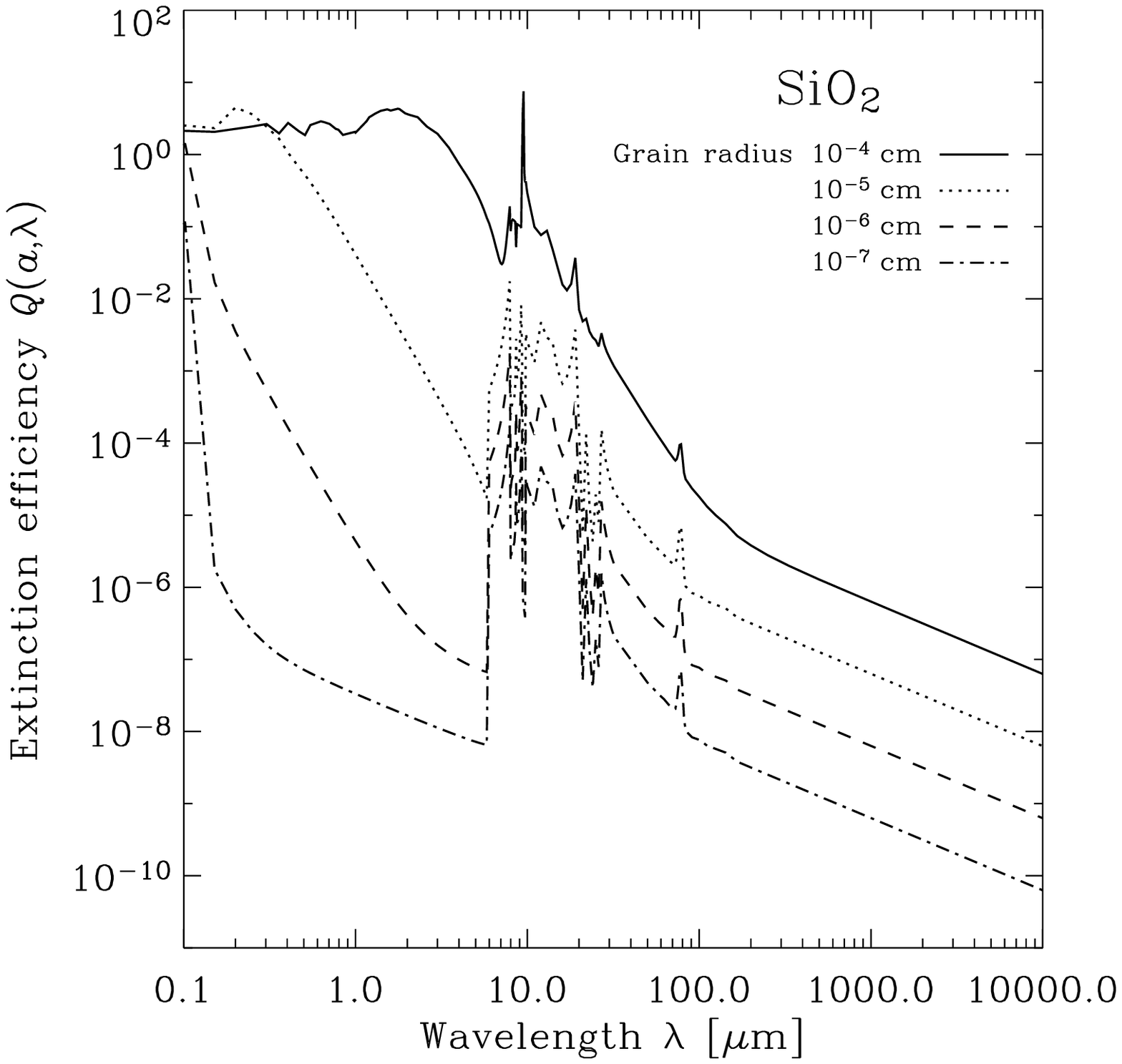}
\centering\includegraphics[width=0.3\textwidth]{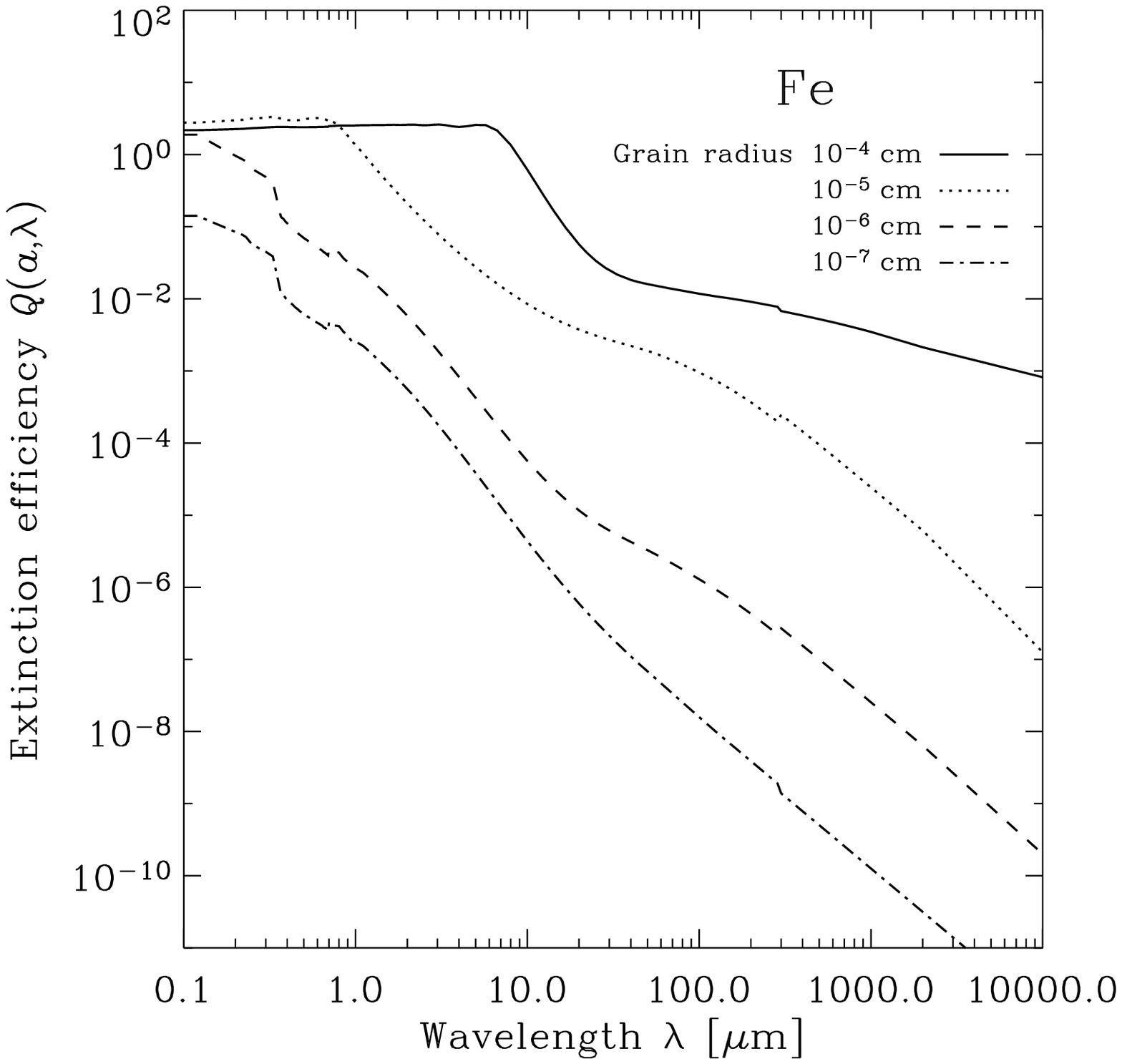}
\centering\includegraphics[width=0.3\textwidth]{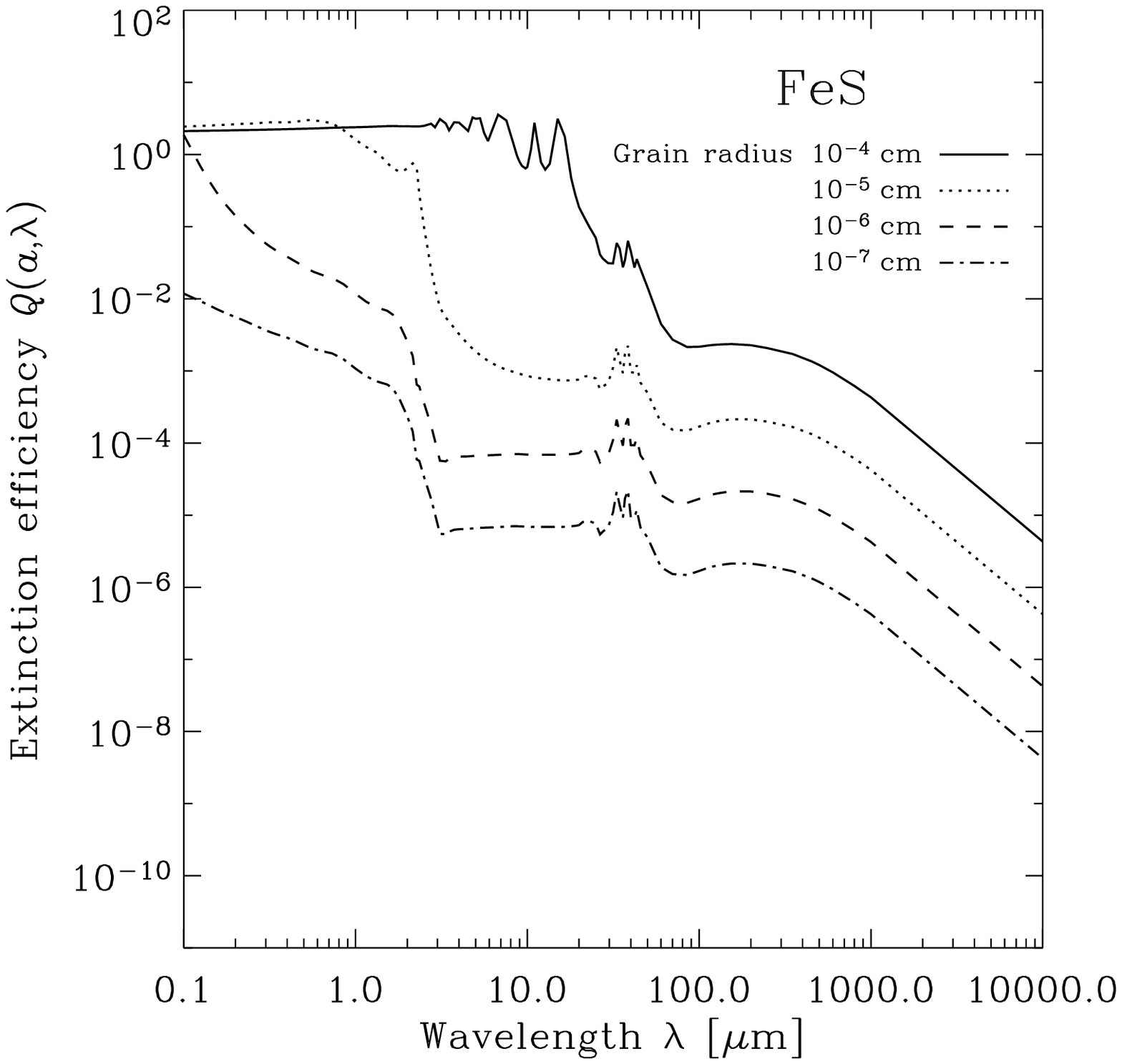}
\centering\includegraphics[width=0.3\textwidth]{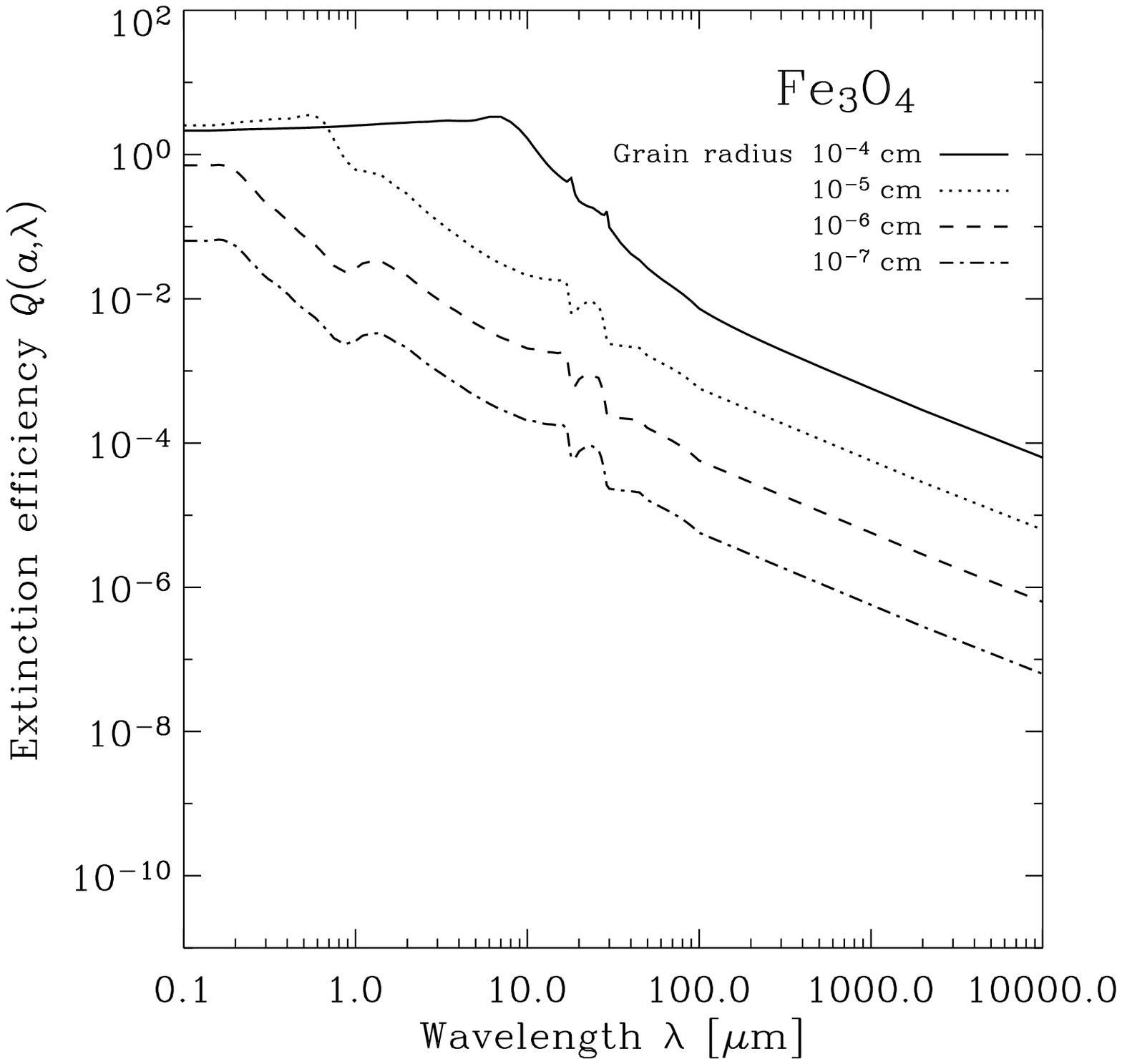}
\centering\includegraphics[width=0.3\textwidth]{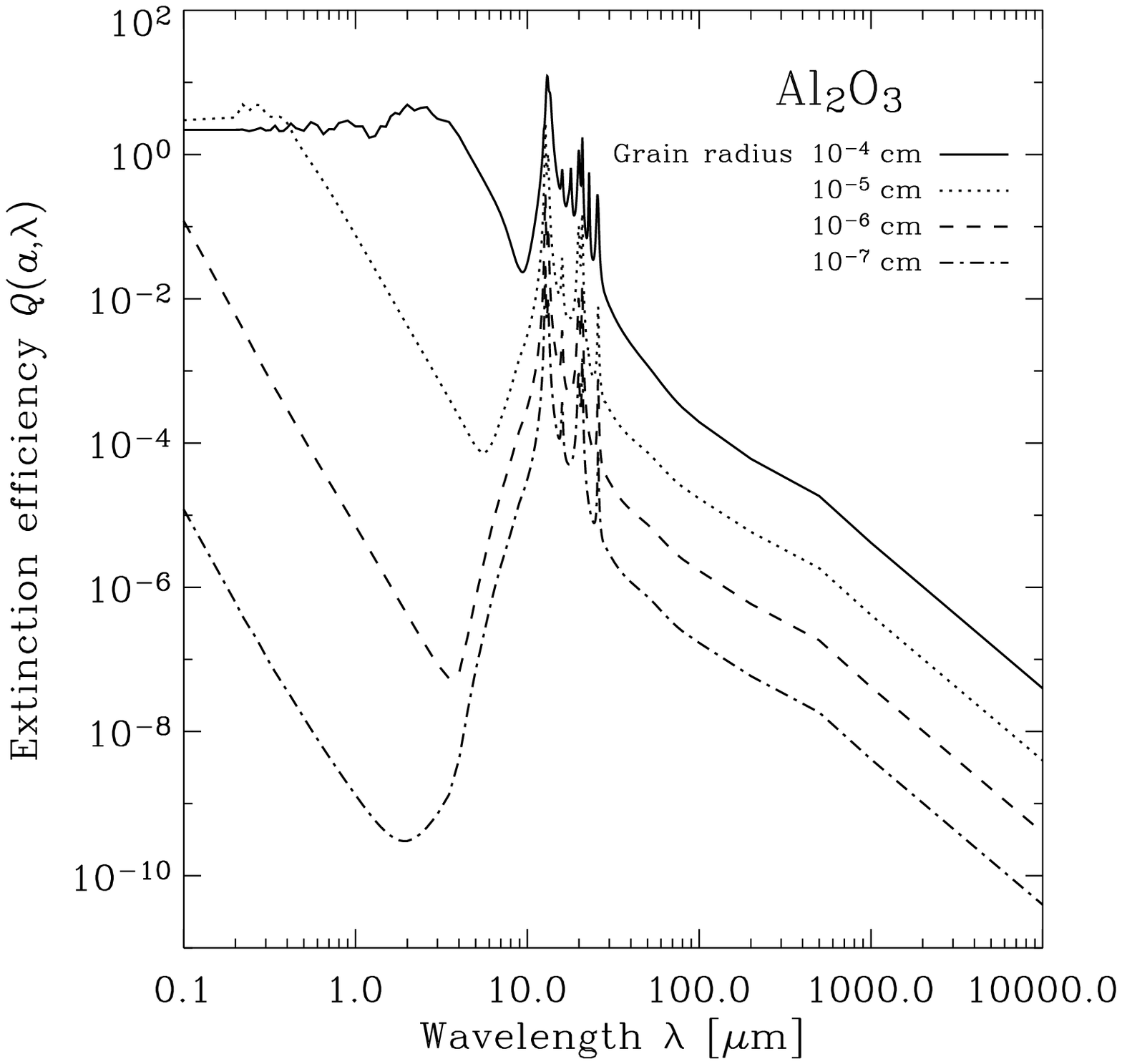}
\centering\includegraphics[width=0.3\textwidth]{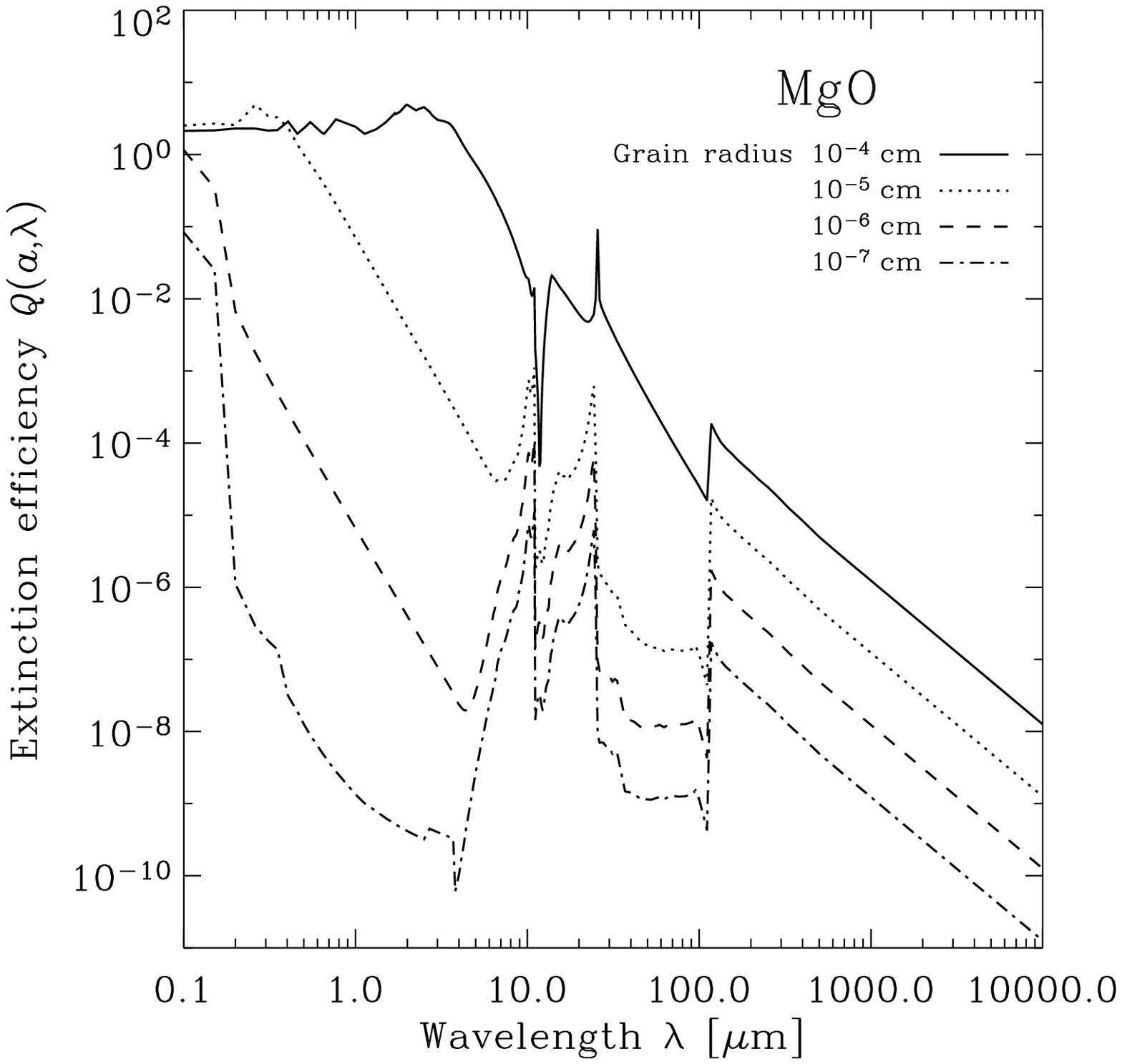}
\centering\includegraphics[width=0.3\textwidth]{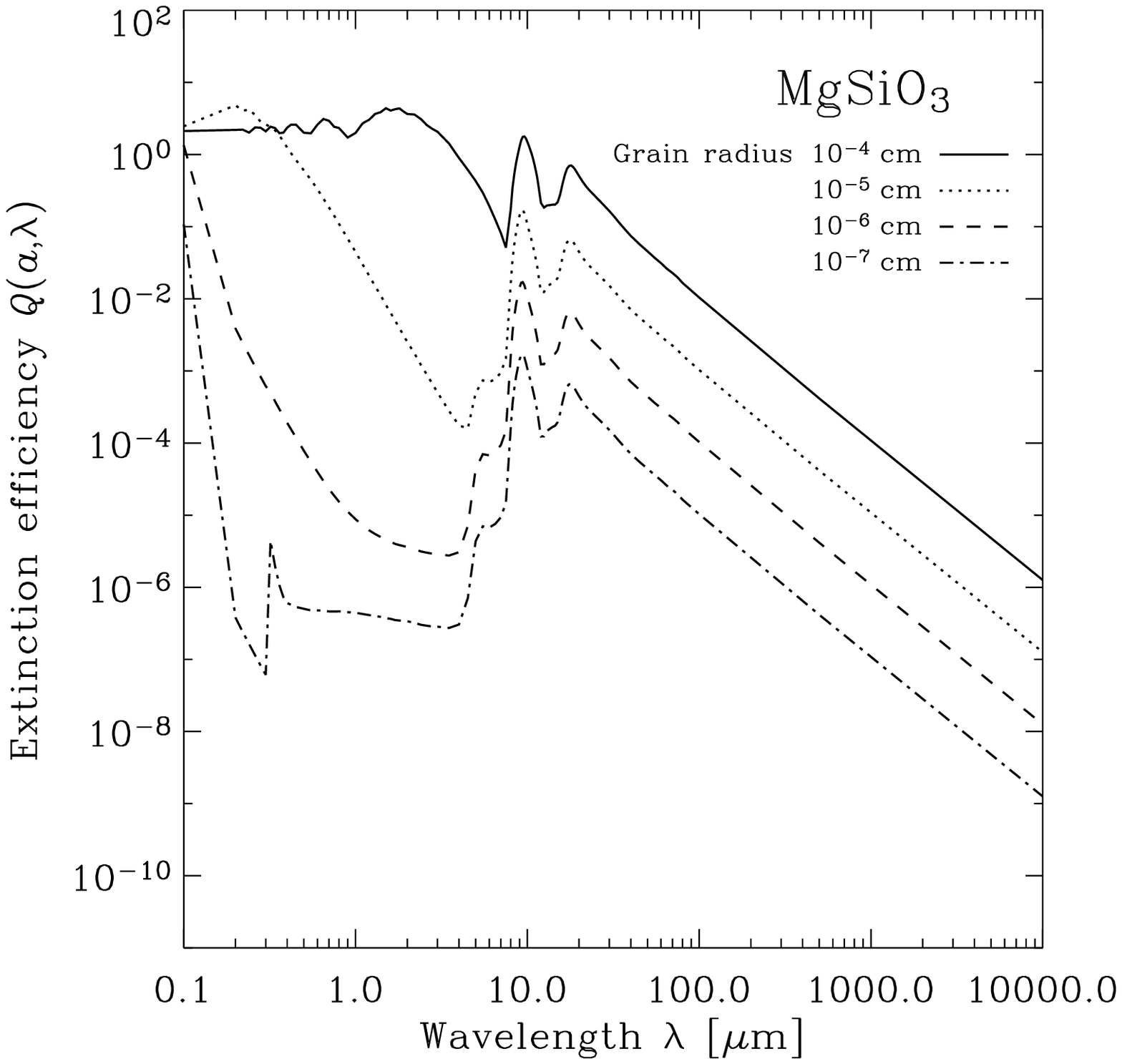}
\centering\includegraphics[width=0.3\textwidth]{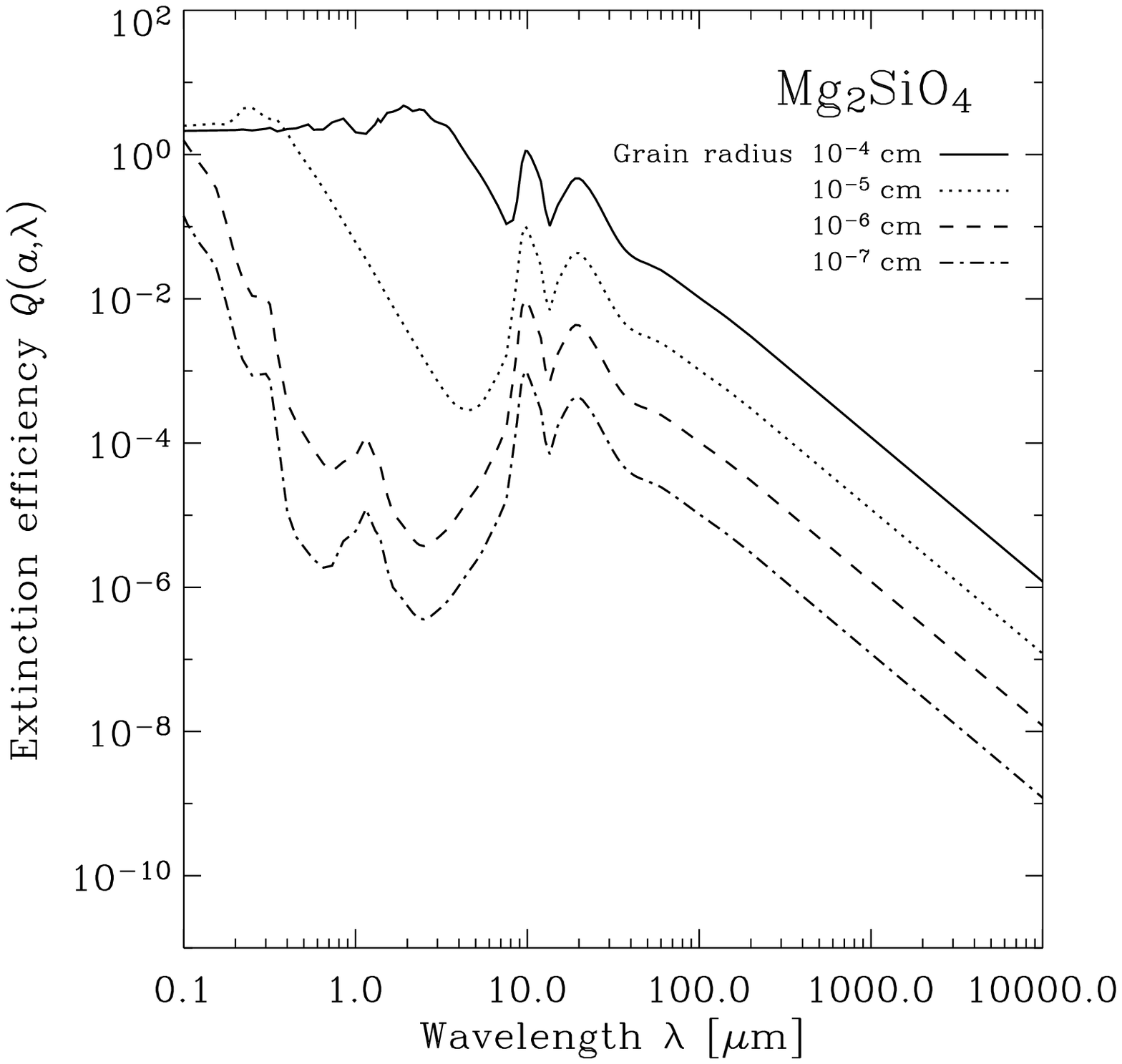}
\caption{
Extinction efficiency $Q(a,\lambda)$ for the grain species produced by the 
supernova model of \citet{nozawa03}.
We show $Q(a,\lambda)$ for the radius $a=10^{-7}, 10^{-6}, 10^{-5}$ and 
$10^{-4}\,\mbox{cm}$ by the curves from bottom to the top in each panel.
}\label{fig:q_abs}
\end{figure*}

For constructing the chemical evolution model of a young galaxy, 
we adopt the following assumptions:
\begin{enumerate}
\item We use a closed-box model, i.e., we neglect an infall and outflow of gas 
in the scale of a star-forming region.
\item For the initial mass function (IMF), we adopt the Salpeter IMF 
\citep{salpeter55}
\begin{eqnarray}\label{eq:IMF}
  \phi(m) \propto m^{-2.35}
\end{eqnarray}
with mass range of $(m_{\rm l}, m_{\rm u})= (0.1\;M_\odot,100\;M_\odot)$.
\item We neglect the contribution of SNe Ia and winds from low-mass 
evolved stars to the formation of dust, because we consider the timescale 
younger than $10^9\;\mbox{yr}$.
\item The interstellar medium is treated as one zone, and the growth of 
dust grains by accretion is neglected.
Within the short timescale considered here, it can be assumed safely
\citep[see, e.g.,][pp.223--224]{whittet92}.
\item We also neglect the destruction of dust grains within the young age
considered \citep[see e.g.,][]{jones96}.
\item We assumed a constant SFR for simplicity.
\end{enumerate}
using these assumptions, we calculate the chemical evolution.
Details of the formulation are presented in 
Appendix~\ref{sec:chemical_evolution}.

The evolution of the total dust amount for $\mbox{SFR} =
1\;M_\odot\,\mbox{yr}^{-1}$ is shown in Figure~\ref{fig:dust_amount}.
The dust mass fraction for unmixed and mixed cases are summarized in
Tables~\ref{tab:mass_fraction_unmixed} and \ref{tab:mass_fraction_mixed},
respectively.
Comparing the evolution of dust mass given by T03, dust mass starts to 
accumulate later than the case of T03, and gradually approaches the T03 
result toward the age of about $10^8\;$Gyr.
This difference is caused by the different formulae we adopted
in the calculations of stellar lifetime: the formula of 
\citet{schaerer02} gives a longer lifetime for the same stellar mass 
than that of \citet{inoue00} which is used in T03.
At $10^8\;$Gyr, T03 and present work yield the same dust mass.

\begin{table}
\begin{center}
\begin{minipage}{0.45\textwidth}
\caption{The specific heat of each species.}
\begin{tabular}{@{}lcccl@{}}\hline
Species	      & Debye temperature & Ref\footnote{
References for Debye temperatures:
(1) \citet{draine01},
(2) \citet{oganov00},
(3) \citet{siethoff96},
(4) \citet{gronvold91},
(5) \citet{shepherd91},
(6) \citet{mezzasalma00},
(7) \citet{hama99}.} 
                         & $N_{\rm atom}$            
                                & $C(T)$ model\footnote{
`Debye' means that the specific heat $C(T)$ is modeled by the classical
Debye model, while `Multi-Debye' means that the multidimensional
Debye model proposed by \citet{draine01} is adopted.
For the species to which we adopt the 
multidimensional Debye model, we do not give the classical Debye temperature.}
\\
              & [K]  &   & [$10^{22}\mbox{cm}^{-3}$] &  \\ \hline
C             & 863  & 1 & 11.4 & Multi-Debye           \\
Si            & 616  & 2 & 5.00 & Debye                 \\
SiO$_2$       & ---  & 1 & 7.96 & Multi-Debye           \\
Fe            & 500  & 3 & 8.50 & Debye                 \\
FeS           & 384  & 4 & 6.62 & Debye                 \\
Fe$_3$O$_4$   & 511  & 5 & 28.4 & Debye                 \\
Al$_2$O$_3$   & 1030 & 6 & 23.5 & Debye                 \\
MgO	      & 762  & 7 & 10.7 & Debye                 \\
MgSiO$_3$     & ---  & 1 & 9.57 & Multi-Debye           \\
Mg$_2$SiO$_4$ & ---  & 1 & 9.63 & Multi-Debye           \\ \hline
\end{tabular}
\label{tab:dust_debye}
\end{minipage}
\end{center}
\end{table}

\subsection{SED construction}\label{subsec:sed_construction}

In this subsection, we present the construction of the SED from dust.
Since most of the formulations are in parallel to those in T03 and T04, 
here we list the essence of our calculation.

\begin{enumerate}
\item{\sl Stochastic heating of very small grains}

Very small grains cannot establish thermal equilibrium with the ambient 
radiation field, which is called stochastic heating \citep[e.g.,][]{krugel03}.
{}To treat this effect, we applied the Debye model to the specific 
heat of the grain species as discussed in T03 and T04.
We adopt a multidimensional Debye model \citep[e.g.,][]{draine01}(DL01) 
for carbon (C) and silicate (SiO$_2$, MgSiO$_3$, and Mg$_2$SiO$_4$) grains.
For other species, we adopt the classical three-dimensional Debye model
with a single Debye temperature.
The specific heat model is summarized in Table~\ref{tab:dust_debye}.

\item{\sl Emission}

The emission from dust is calculated in the same way as T03/T04, 
basically according to \citet{draine85}.
Total dust emission is obtained as a superposition of the emission from 
each grain species.
The total mass of each grain component is given by N03.
With this value and material density of each species
(Table~\ref{tab:dust_species}), we can determine the normalization of
the dust size distribution (see Appendix~\ref{sec:normalization}).

We constructed $Q(a, \lambda)$ of each grain species from available 
experimental data via Mie theory \citep[e.g.,][Chap.~2]{krugel03}.
The extinction efficiencies are presented in Figure~\ref{fig:q_abs}.
T03 adopted \citet{draine84} for the optical properties of these species,
being different from the present work.
On the contrary, we treat detailed species in the present work.
Hence, the mass fraction distributed to carbon and silicate are significantly
reduced comparing to the case of T03.
In the IR, $Q(a,\lambda)$ of these grains is much larger than other grains 
at the whole range of grain size.
As a result, the sum of the contributions of all grains are less than that
of the case where only two species are considered as T03.

\item{\sl Extinction}

Self-absorption in the MIR for a very optically thick case is treated 
by a thin shell approximation, in the same manner as T04.
\end{enumerate}

\section{Results}\label{sec:results}

\begin{figure*}
\centering\includegraphics[angle=90,width=0.9\textwidth]{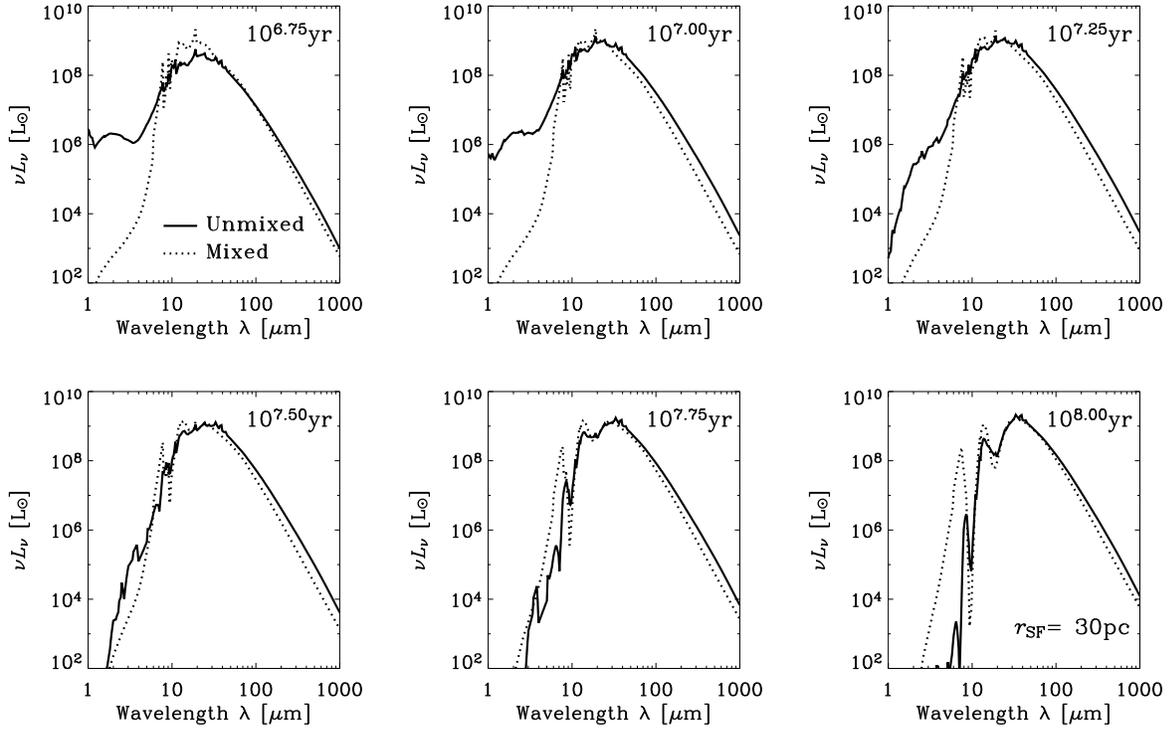}
\caption{
The evolution of the infrared (IR) spectral energy distribution 
(SED) of a very young galaxy.
The size of the star forming region, $r_{\rm SF}$, is 30~pc.
Solid and dotted lines represent the SEDs based on the unmixed and mixed 
model of dust production \citep{nozawa03}, respectively.
For Figures~\ref{fig:sed_30pc} and \ref{fig:sed_100pc}, a constant 
star formation rate of ${\rm SFR}=1\,M_\odot \mbox{yr}^{-1}$ is adopted.
}\label{fig:sed_30pc}
\end{figure*}

\begin{figure*}
\centering\includegraphics[angle=90,width=0.9\textwidth]{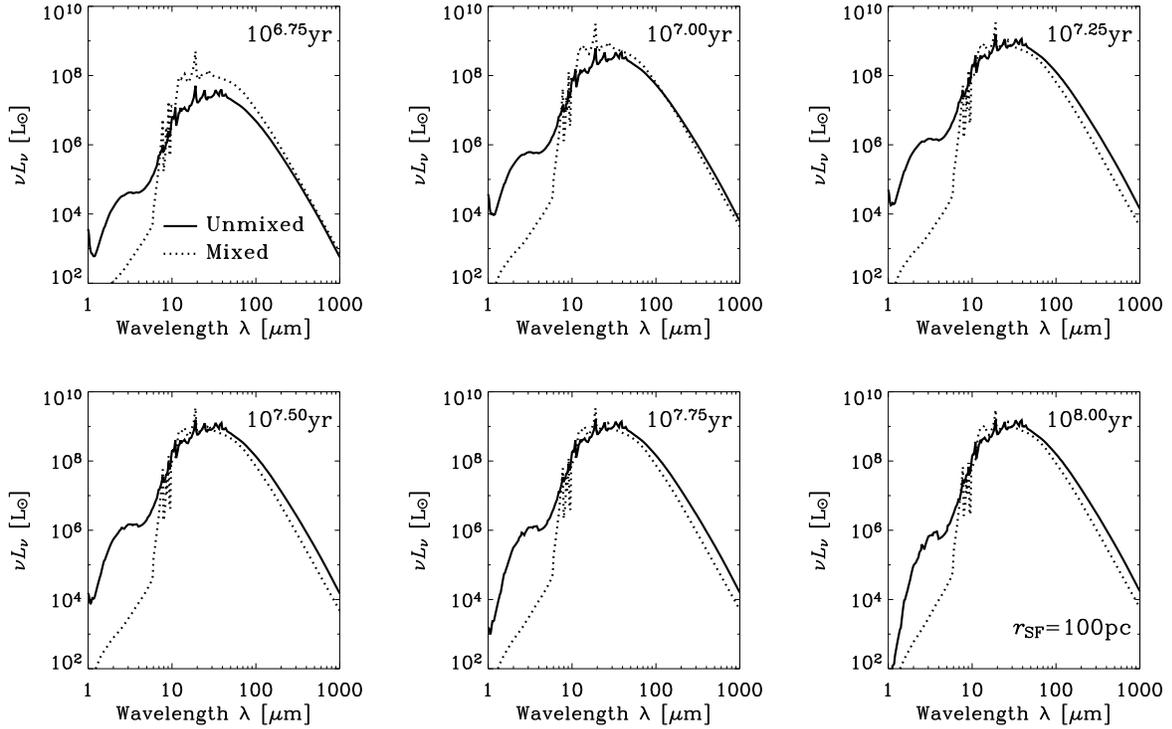}
\caption{
Same as Figure~\ref{fig:sed_30pc}, but the size of the star forming region
$r_{\rm SF}$ is $100\,\mbox{pc}$.
}\label{fig:sed_100pc}
\end{figure*}

\begin{figure*}
\centering\includegraphics[width=0.45\textwidth]{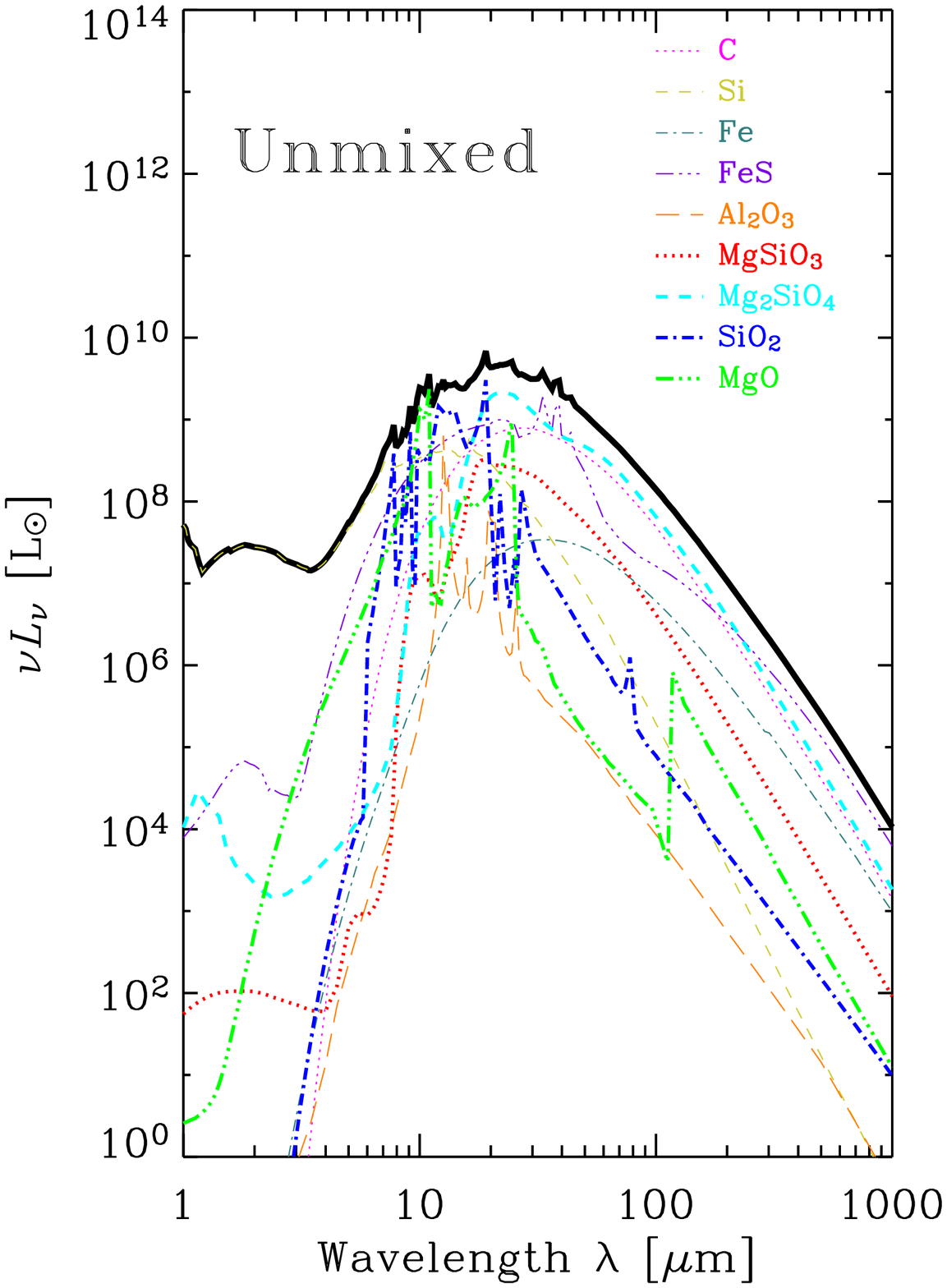}
\centering\includegraphics[width=0.45\textwidth]{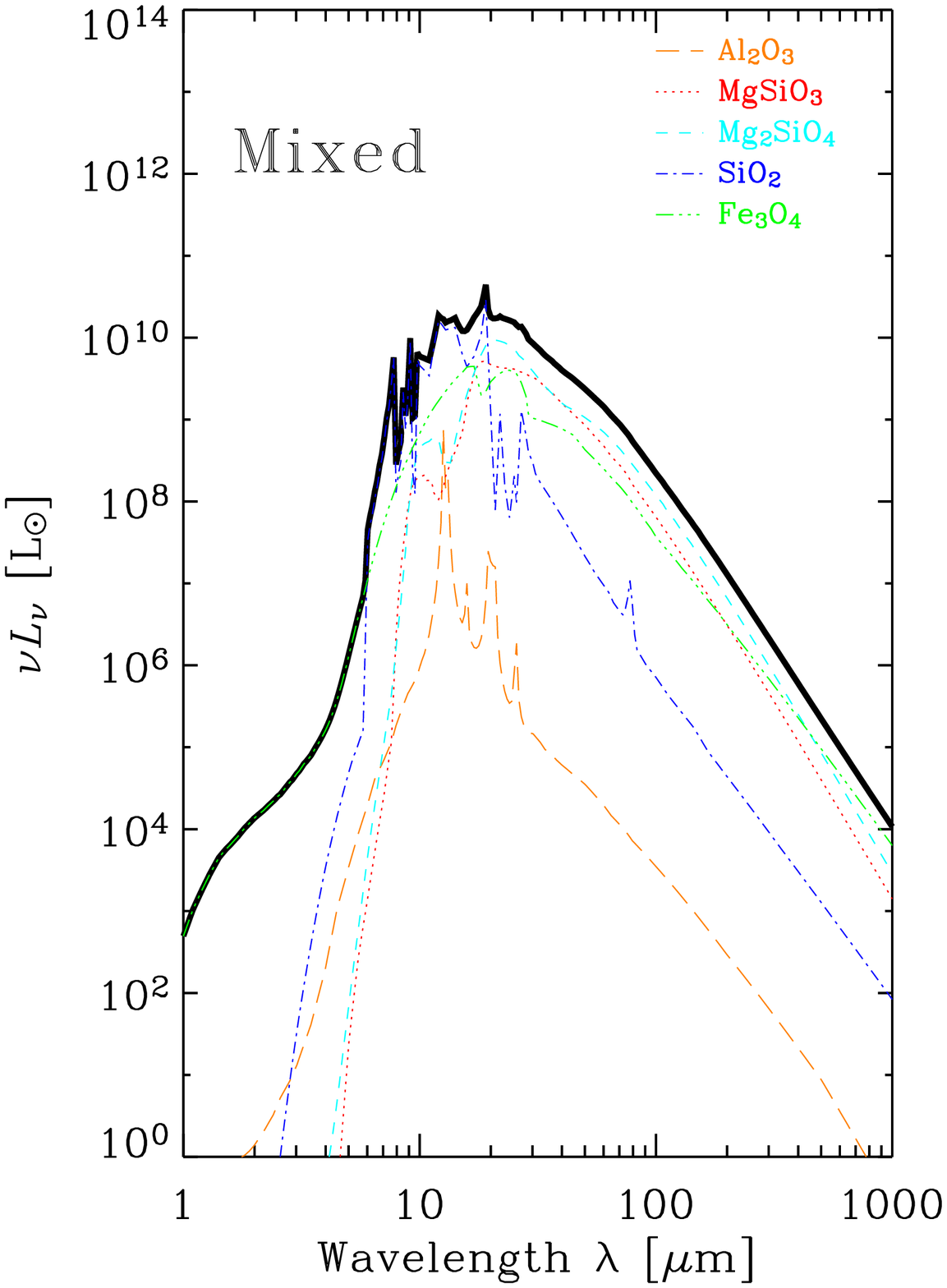}
\caption{The contribution of each dust species to the total SED of a young
galaxy.
In this figure we set the galaxy burst age to be $t=10^{6.75}\,\mbox{yr}$
and the radius of the star forming region to be $r_{\rm SF} = 30\,\mbox{pc}$.
Left panel is the SED calculated from the unmixed dust production,
and right panel is obtained from the mixed dust production.
}\label{fig:sed_contribution}
\end{figure*}

\subsection{Evolution of infrared SED}

We first show the evolution of the IR SED of forming galaxies based on 
our baseline model.
For these calculation we adopted the star formation rate ${\rm SFR}=1\,M_\odot
\,\mbox{yr}^{-1}$.
We adopt $r_{\rm SF} = 30$~pc and 100~pc, the same as those used in T03.
These values are relevant when describing `dwarf-like' young galaxies. 
The former successfully reproduce various properties of active star-forming
dwarf galaxies like \sbs, whereas the latter is representative of 
the size of quiescent dwarf galaxies like \izw\ \citep{hirashita04}.
We will revisit these representative dwarf starburst galaxies in 
Section~\ref{sec:discussion}.

The results are presented in 
Figures~\ref{fig:sed_30pc} and \ref{fig:sed_100pc}.
Figure~\ref{fig:sed_30pc} is the SED of a galaxy with 
$r_{\rm SF}=30\,\mbox{pc}$, while Figure~\ref{fig:sed_100pc} is the one for
$r_{\rm SF}=100\,\mbox{pc}$.
We calculated the evolution of the SED in the age range of 
$10^{6.5}\mbox{--}10^{8}$~yr.
In a very young phase ($\mbox{age}=10^{6.75}\mbox{--}10^{7.25}$~yr), 
unmixed-case SED has an enhanced N--MIR continuum.
After $10^{7.25}$~yr, the N--MIR continuum is extinguished by the 
self-absorption in the case of $r_{\rm SF}=30$~pc. 
In contrast, the self-absorption is not significant for $r_{\rm SF}=100$~pc.

In both cases, the SEDs have their peaks at a wavelength $\lambda \simeq 
20\mbox{--}30\;\mu$m, which is much shorter than those of dusty giant galaxies
at $z=1\mbox{--}3$ detected by SCUBA.
At submillimetre wavelengths, the shape of the continuum is very 
similar to each other for mixed and unmixed cases, though the unmixed case
predicts stronger fluxes by a factor of two.
This is explained by the grain size distributions of the unmixed and mixed 
cases:
carefully examining Figure~\ref{fig:dust_size}, we see that the largest-size
dust grains are more abundant for the unmixed-case size distribution.
It means that larger grains contribute more significantly than smaller ones.
Since larger grains radiate their energy at longer wavelengths, the total SED
has a stronger FIR-submm continuum.

As a whole, our model SEDs are less luminous than those of T03 by a factor
of 2--3.
This is due to the difference in considered grain species, 
and more importantly, also due to the improvement of the treatment of 
$Q(a,\lambda)$ at the UV;
$Q(a,\lambda)$ was approximated to be unity in T03.
Also T03 treated the size dependence of $Q(a,\lambda)$ by a simple scaling law,
which was not very accurate.
As seen in Figure~\ref{fig:q_abs}, the size dependence of $Q(a, \lambda)$
cannot be described by a simple scaling law, especially at the MIR
(see, e.g., Si, Fe, and FeS).
We, in this work, used the exact value of $Q(a, \lambda)$ in all the 
wavelengths.
Since some elements have a very small value for $Q(a,\lambda)$ at the UV 
regime, the total absorption probability of UV photons is smaller than
that of T03.
A strong N--MIR continuum at a very young phase and the dependence on 
$r_{\rm SF}$ are qualitatively consistent with the previous result of T03.
However, the dust grains expected from N03 model predict a peak of the 
SED at shorter wavelengths than those of T03.
In addition, a relatively small amount of small dust grains also makes 
the N--MIR SED weaker than the T03 one.

\subsection{Contribution of each species to the SED}

The dust species treated here are much more detailed than those of T03.
In order to see the detailed contributors to the total SEDs, we show the 
individual SED of each species in Figure~\ref{fig:sed_contribution}.
We show the SED with the age of burst $10^{6.75}$~yr, where the effect of 
self-absorption is negligible.

Both in unmixed and mixed cases, the smallest dust grain species is 
Al$_2$O$_3$ (Figure~\ref{fig:dust_size}).
However, because of its small fraction in mass (Figure~\ref{fig:dust_amount}), 
Al$_2$O$_3$ contributes to the total SED very little.
The main contributor to the (unextinguished) N--MIR continuum is Si for unmixed
case and Fe$_3$O$_4$ in mixed case, respectively.
These species are the second smallest grains.
For the unmixed case, since the emissivity of Si grains with the size of 
$\sim 10^{-7}\;\mbox{cm}$ is much smaller than that of Fe$_3$O$_4$ at 
$\lambda \ga 1\;\mu$m, the grain temperature becomes much higher than that of
Fe$_3$O$_4$, which makes the very strong N--MIR continuum emission.
On the contrary, since $Q(a,\lambda)$ of Fe$_3$O$_4$ is large, 
the resultant SED for the mixed case does not have a strong N--MIR continuum.

At $\lambda=10\mbox{--}20\;\mu$m, several species contribute equally to the
SED in unmixed case, while SiO$_2$ play a dominant role for mixed case.
At longer wavelengths, the main contributors to the SED are
Mg$_2$SiO$_4$ and amorphous carbon for the unmixed case, and 
MgSiO$_3$ and Mg$_2$SiO$_4$ in mixed case.
In addition, especially at $\lambda \ga 200\;\mu$m, FeS (unmixed) and
Fe$_3$O$_4$ (mixed) are also important because of their shallow slope of 
the extinction efficiency at longer wavelength regime.
Note that, for the mixed case, Fe$_3$O$_4$ contribute to both NIR and 
FIR.

\subsection{Opacity and its evolution}

\begin{figure*}
\centering\includegraphics[angle=90,width=0.9\textwidth]{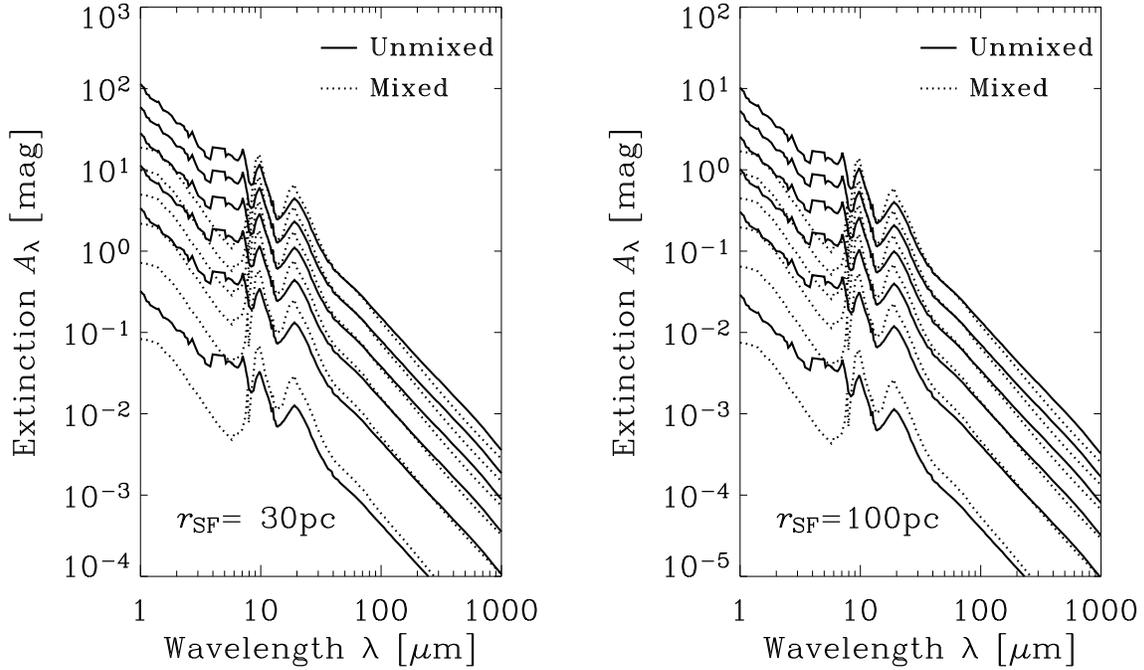}
\caption{
The evolution of the opacity of a very young galaxy at IR wavelengths.
Left panel shows the IR extinction curves calculated with the star forming 
region size $r_{\rm SF} = 30\,\mbox{pc}$, and right panel shows the curves 
with $r_{\rm SF} = 100\,\mbox{pc}$.
{}From the bottom to the top, the ages of galaxies are
$10^{6.75}\,\mbox{yr}$,
$10^{7.0}\,\mbox{yr}$,
$10^{7.25}\,\mbox{yr}$,
$10^{7.5}\,\mbox{yr}$,
$10^{7.75}\,\mbox{yr}$,
and 
$10^{8.0}\,\mbox{yr}$, respectively.
The solid and dashed curves represent the IR extinction curves based on 
the unmixed and mixed model, respectively.
}\label{fig:ext}
\end{figure*}

The extinction curve $A_\lambda$ is obtained as
\begin{eqnarray}
  A_\lambda = 1.086\tau_{\rm dust}(\lambda)\;.
\end{eqnarray}
In this work we concentrate on the extinction curve in the IR.
Detailed discussions of the extinction curves in the optical wavelength
regime are given in \citet{hirashita05}.\footnote{
Since we assume a very simple geometry, the extinction curve is directly
related to the `attenuation curve', in which the geometrical effect is 
also included \citep[for a thorough discussion on this subject, 
see, e.g., ][]{inoue05}.}

Figure~\ref{fig:ext} presents the evolution of the IR opacity for a young
galaxy with the age of $10^{6.75}\mbox{--}10^{8}\;$yr (from the bottom to 
the top).
For the case of $r_{\rm SF}=30$~pc, the extinction becomes almost unity at 
$10\;\mu$m when the age is $10^{7.25}$~yr.
Hence, after this age, the N--MIR regime of the SED becomes optically thick,
and this makes the continuum extinguished in Figure~\ref{fig:sed_30pc}.
On the other hand, for $r_{\rm SF}=100$~pc, it remains optically thin even
at the age of $10^8$~yr at $10\;\mu$m.

In Figure~\ref{fig:ext}, we find an interesting difference between the
extinction curves of the unmixed and mixed cases.
Though their behavior is similar to each other at wavelengths 
$\lambda \ga 10\;\mu$m, there is no dip at $\lambda  \la 10\;\mu$m
in the curve of the unmixed case.
This makes the extinction stronger for the unmixed-case galaxies at N--MIR.
In addition, the MIR bumps at 10 and $20\;\mu$m are less prominent for 
unmixed case.

Comparing these IR extinction curves with T03 model, we find that
the amount of the extinction is smaller than that of T03 by a factor of 2--10,
depending on the wavelength and unmixed/mixed production.
This is explained by the discussion presented in 
Subsection~\ref{subsec:sed_construction}: T03 supposed silicates and carbon
grains as the constituents of dust grains from SNe. Since their $Q(a,\lambda)$
is larger than the other species predicted by N03, the total extinction 
becomes smaller in this work.

As for the shape of the curve, the extinction curve for the mixed model 
resembles that of T03 extinction curves, while the curve for the unmixed
model is qualitatively different at MIR, because of the weak MIR bump features
and the lack of the MIR dip mentioned above.

\begin{figure*}
\centering\includegraphics[width=0.45\textwidth]{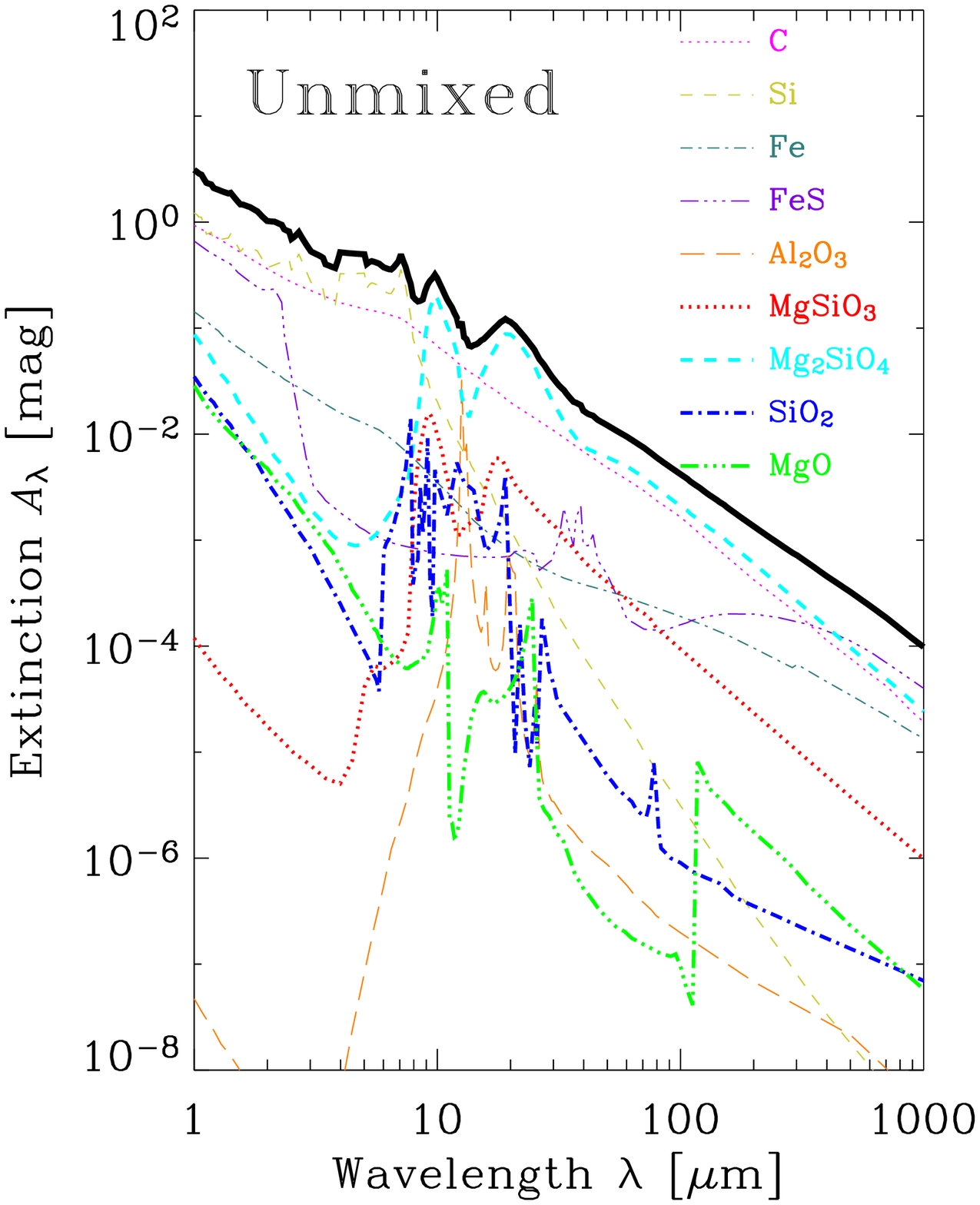}
\centering\includegraphics[width=0.45\textwidth]{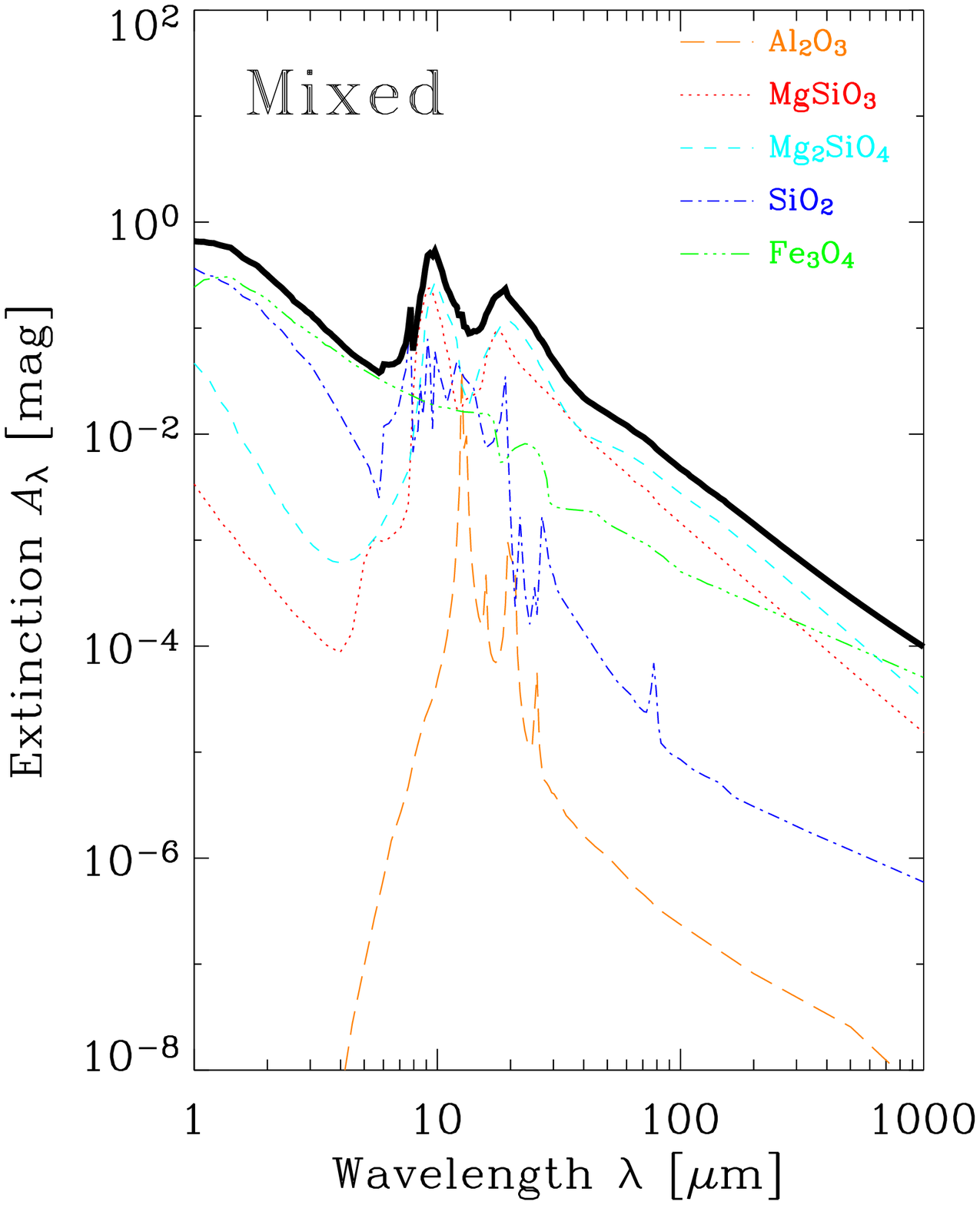}
\caption{The contribution of each dust species to the total IR extinction
curve of a young galaxy.
In this figure we set the galaxy burst age of $t=10^7\,\mbox{yr}$.
Left panel is the extinction curve calculated from the unmixed dust production,
and right panel is obtained from the mixed dust production.
}\label{fig:tau_contribution}
\end{figure*}

Figure~\ref{fig:tau_contribution} demonstrates the contribution of each 
dust grain species to the total IR extinction curves.
The adopted age is $10^7$~yr as a representative, but the contribution is 
constant with time.

For the extinction curve of the unmixed case, bump features at MIR are
relatively weak (Figure~\ref{fig:ext}).
Similar to those of the Galactic IR extinction curve, they are due to silicate 
grains.
In the unmixed-model extinction, the bumps are dominated by Mg$_2$SiO$_4$, 
and the contribution from MgSiO$_3$ is not important.
We also find a strong contribution from amorphous carbon grains which have
very smooth dependence of $Q(a,\lambda)$ on wavelength.
They makes the silicate features rather weak, because of their comparable
contribution with Mg$_2$SiO$_4$.
We note that the amorphous carbon does not have any dip at $\lambda \simeq 
5\mbox{--}10\;\mu$m.
Further, large grains of Si have a large $Q(a,\lambda)$ at these wavelength
regime.
We find that both the amorphous carbon and Si grains equally contribute to
the total extinction at N--MIR, and these species plug up the MIR dip of 
Mg$_2$SiO$_4$.
At NIR wavelengths close to $1\,\mu$m, FeS also contributes to the total
extinction.

We next see the mixed case.
In this case, MgSiO$_3$ and Mg$_2$SiO$_4$ equally contribute to the total
extinction around the MIR bumps.
In contrast to the unmixed case, there is no species which fill up the dip of 
silicates, hence we see clear silicate bump features and dip in the IR 
extinction curve.
In the N--MIR, main contributors to the opacity are Fe$_3$O$_4$ and SiO$_2$.
The contribution from Al$_2$O$_3$ is not important at any wavelength.

\section{Discussion}\label{sec:discussion}

\subsection{Nearby forming dwarf galaxies}

It is still a difficult task to observe galaxies in their very first phase 
of the SF, especially to detect their dust emission directly.
Along the line of studies made by \citet{hirashita02a}, T03 considered two 
local star-forming dwarf galaxies, \sbs\ and \izw, for investigating the
dust emission from very young small galaxies.
Here we revisit these representative star-forming dwarfs with our framework.
Since a recent observation of \sbs\ by {\sl Spitzer} has
been reported \citep{houck04}, it is timely to reconsider these 
`textbook objects' with the new data.
In addition, understanding the SEDs of these objects will shed light to 
the physics of interstellar matter and radiation of high-$z$ galaxies also 
via empirical studies 
\citep[e.g.,][]{takeuchi03b,takeuchi05}.\footnote{We should keep in 
mind that the following discussion is limited to a particular class of dwarf
galaxies dominated by newly formed stars.
For the modeling of the SED of normal star-forming dwarfs, see, e.g., 
\citet[][]{galliano03,galliano05}.
}

\subsubsection{\sbs}

\begin{figure*}
\centering\includegraphics[width=0.45\textwidth]{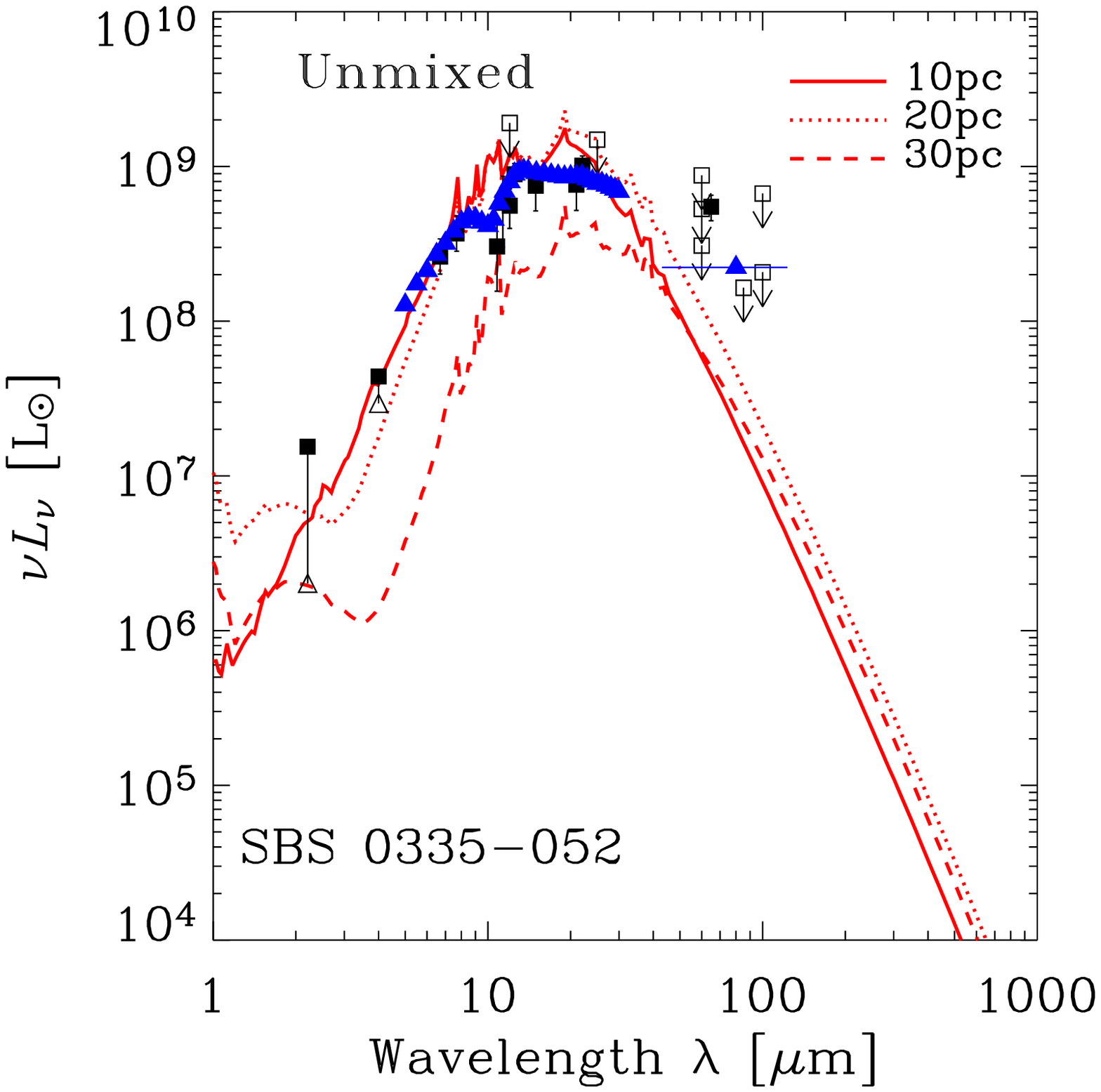}
\centering\includegraphics[width=0.45\textwidth]{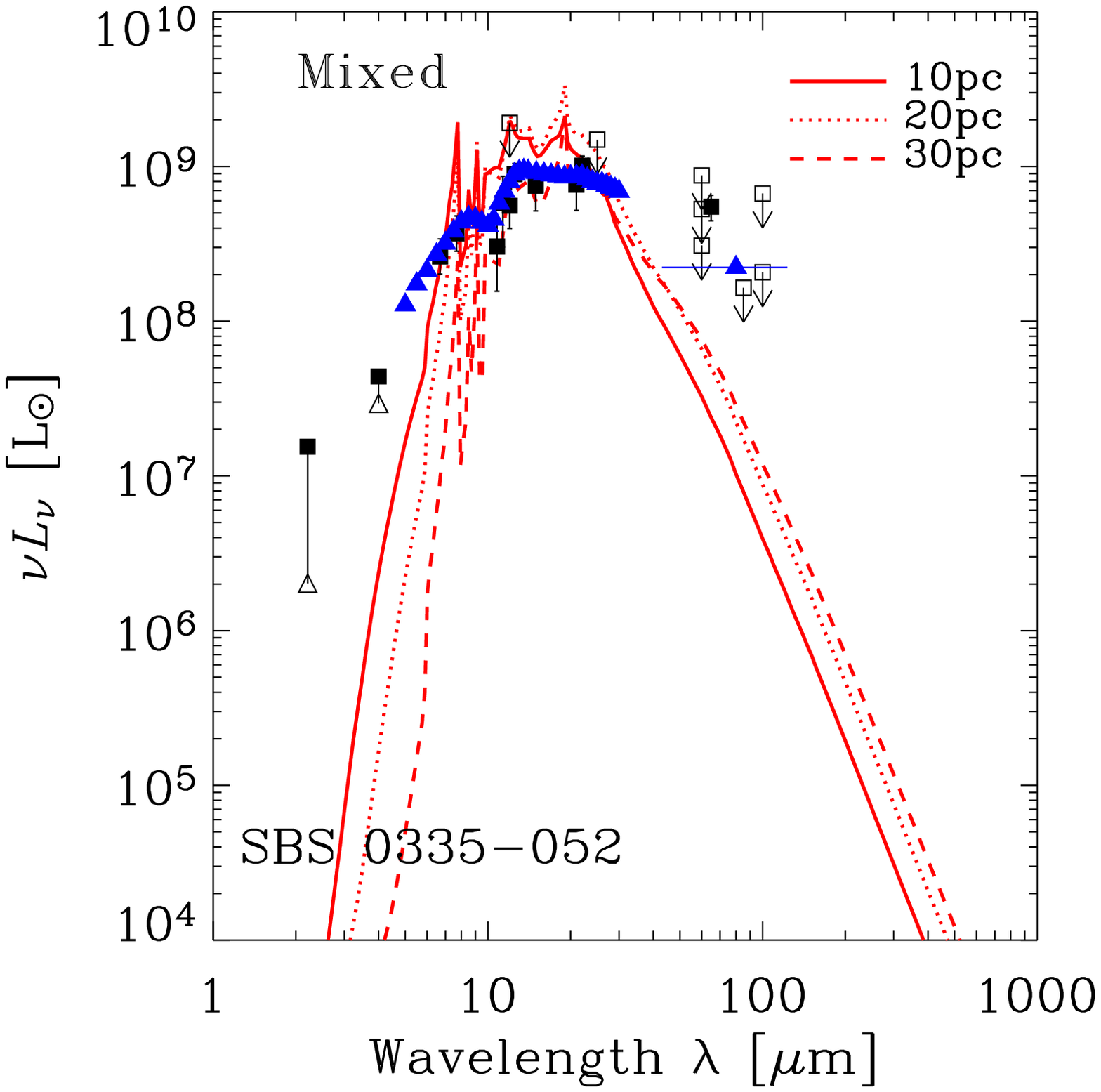}
\caption{The model for the SED of a nearby star-forming dwarf galaxy \sbs.
We adopt the star formation rate $\mbox{SFR}=1.7\;M_\odot\,\mbox{yr}^{-1}$ 
\citep{hunt01}.
The galaxy burst age is found to be $t=10^{6.5}\,\mbox{yr}$ and the
radius of the star forming region is set to be $r_{\rm SF} = 10$, 20, and
30~pc.
Solid squares are the observed data from {\sl ISO}, while open squares 
represent the upper limits obtained from {\sl IRAS} and {\sl ISO} observations.
Open triangles are the expected contribution of dust emission calculated by 
the recipe of \citet{joy88}.
Solid triangles are taken from recent observation of \sbs\ by {\sl Spitzer}
\citep{houck04}.
Left panel is the SED calculated from the unmixed dust production,
and right panel is obtained from the mixed dust production.
Clearly, the mixed dust production poorly reproduces the observed SED of \sbs.
}\label{fig:sed_sbs}
\end{figure*}

\sbs\ is a local galaxy ($\sim 54\;\mbox{Mpc}$) with $\mbox{SFR} = 1.7 
\;M_\odot\,\mbox{yr}^{-1}$ \citep{hunt01} and extremely low metallicity
$Z = 1/41\,Z_\odot$.
This galaxy is known to have an unusual IR SED and strong flux at N--MIR.
It has a very young starburst ($\mbox{age} \la 5\,\mbox{Myr}$) without 
significant underlying old stellar population \citep{vanzi00}.
T03 have modeled the SED of \sbs\ and reported a good agreement with the
available observations at that time.

However, \citet{houck04} presented new data of the MIR SED by {\sl Spitzer},
and reported a deviation of the model by a factor of two or three.
Their observation indicated that \sbs\ has even more FIR-deficient SED than 
ever thought.
Hence, it is interesting to examine whether our present model can reproduce
the extreme SED of this galaxy.
We are also interested in the possibility to determine which of the two 
pictures, unmixed or mixed, is plausible, by a direct measurement of the SED.

We show the model SEDs for \sbs\ in Figure~\ref{fig:sed_sbs}.
We have calculated the SED for $r_{\rm SF}=10$, 20, and 30~pc both for 
unmixed and mixed cases.
The SFR is fixed to be $1.7\,M_\odot\,\mbox{yr}^{-1}$, and the age is 
$10^{6.5}$~yr.
Solid squares are the observed data from {\sl ISO}, while open squares 
represent the upper limits obtained from {\sl IRAS} and {\sl ISO} observations.
Open triangles are the expected contribution of dust emission calculated by 
the recipe of \citet{joy88}.
Filled triangles depict the measured SED of \sbs\ by {\sl Spitzer} IRS
taken from \citet{houck04} in the MIR.
The filled triangle at FIR is the estimate from the radio observation of 
\citet{hunt04}, which is used by \citet{houck04}.
Details of the other observational data are found in Section~4 of T03.

In the FIR regime, our model SEDs are consistent with the strong constraint 
given by \citet{houck04}, both for the unmixed and mixed cases.
This is because our present model predicts a peak of the SED at shorter 
wavelengths than that of T03.
At MIR, though we cannot give an excellent fit to the observed data, the model
SEDs roughly agree with them, and the unmixed-case SEDs with $r_{\rm SF}=10
\mbox{--}20$~pc give a better fit.
The very strong N--MIR continuum of \sbs\ is well reproduced by the SED
of the unmixed dust production picture.
On the contrary, the SEDs of the mixed case seriously underpredicts the
observationally suggested N--MIR continuum.
Since we can also calculate the extinction value $A_\lambda$, we can 
distinguish between a mixed-case SED and heavily extinguished unmixed SED
without ambiguity.

In summary, the SED for the unmixed dust production with 
$r_{\rm SF}=10\mbox{--}20$~pc yields a reasonable fit to the latest 
observations by {\sl Spitzer}.
This result suggests that we may determine the dust production (unmixed or
mixed) of SNe through the observation of the N--MIR SEDs of forming galaxies.
As for \sbs, the unmixed dust production is suggested to take place.
In a previous work, we have modeled a high-$z$ ($z=6.2$) extinction curve 
\citep{maiolino04} and found that it is well represented by dust produced 
in unmixed SNe \citep{hirashita05}.
This conclusion is strengthened by our result above.

The dust mass calculated by our model at this age of \sbs\ is 
$1\mbox{--}2 \times 10^3\;M_\odot$, consistent with the observationally
estimated value by \citet{dale01}.
We note that the mass estimation is strongly dependent on the assumed
dust species and their emissivities, and grain size distribution.
As we discussed in Section~\ref{sec:results}, the continuum radiation in the
N--MIR is dominated by stochastically heated dust emission, which is
completely different from modified blackbody.
Therefore, when we try to estimate the dust mass, we must take care to 
determine the corresponding grain properties, i.e., radiative processes
and grain which are related to the observed SED of galaxies.

\subsubsection{\izw}

\begin{figure*}
\centering\includegraphics[width=0.45\textwidth]{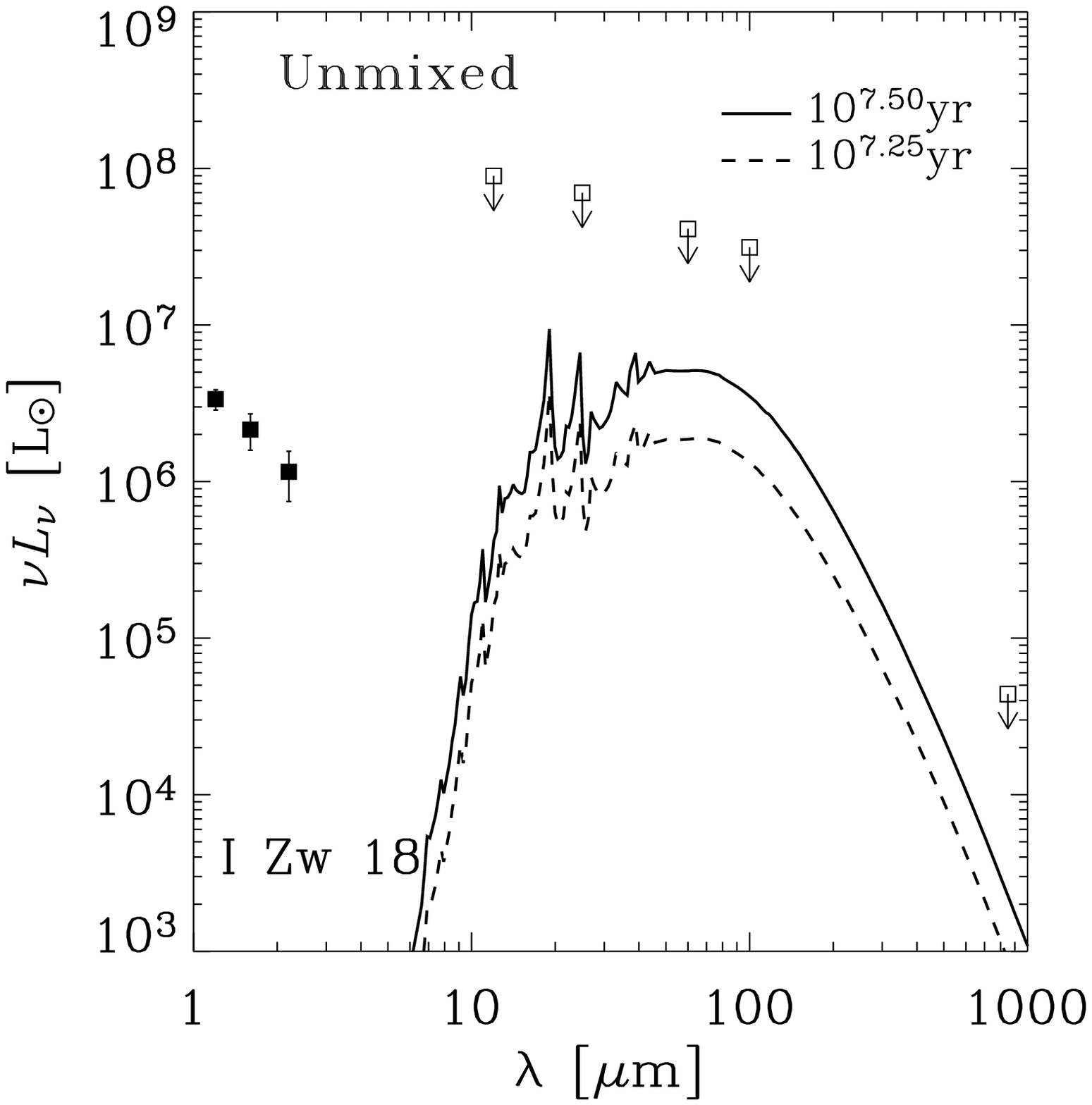}
\centering\includegraphics[width=0.45\textwidth]{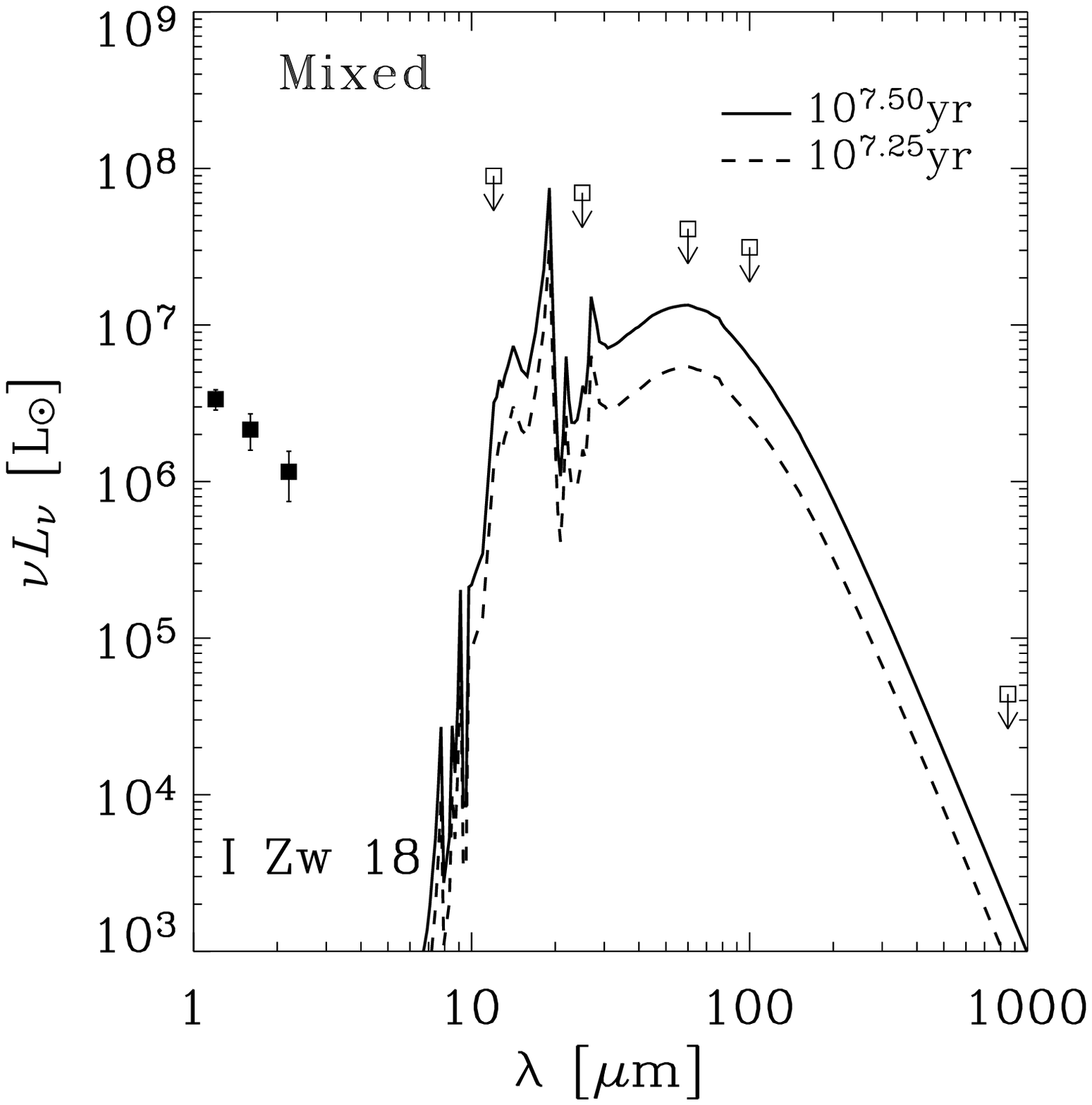}
\caption{A prediction for the SED of a nearby `quiescent' star-forming 
dwarf galaxy \izw.
We adopt $\mbox{SFR}=0.04\;M_\odot\mbox{yr}^{-1}$ and $r_{\rm SF}=100$~pc.
The age of the star formation is set to be $10^{7.25}$~yr (dashed lines)
and $10^{7.5}$~yr (solid lines). 
Since the NIR continuum of \izw\ is expected to be dominated by stellar 
and nebular continuum radiation, the contribution of dust may be negligible,
and it is not necessary to reproduce these NIR observed fluxes by the present
model.
}\label{fig:sed_izw}
\end{figure*}

\izw\ is the most metal-deficient star-forming galaxy in the local universe
($Z=1/50\,Z_\odot$).
We adopt its distance of 12.6~Mpc \citep{ostlin00}, and the SFR of 
$0.04\;M_\odot\,\mbox{yr}^{-1}$.
The existence of the underlying old stellar population is suggested for
this galaxy, but their contribution is not dominant \citep{hunt03}.
Thus, it is not unreasonable to adopt our present model to predict the
IR SED of \izw.

\citet{cannon02} estimated the total dust mass $M_{\rm dust} = 2\mbox{--}3
\times 10^3\,M_\odot$ from {\sl HST} WFPC2 narrow-band imaging.
In our model, the dust mass reaches this value at burst age of 
$10^{7.25}\mbox{--}10^{7.5}$~yr.
This age is consistent with the observationally suggested age of its major SF
\citep[e.g.,][]{hunt03}.
\citet{cannon02} obtained the extinction of $A_V=0.5$~mag for some patches 
in this galaxy, which is reproduced by our model with the above age.

We show our model prediction for the SED of \izw\ in Figure~\ref{fig:sed_izw}.
We present the SED for the age $10^{7.25}$~yr and $10^{7.5}$~yr.
The NIR flux measurements are taken from \citet{hunt03}.
\izw\ is not detected at all the four {\sl IRAS} bands, hence
we calculated the upper limits in the M--FIR from the sensitivity limits of 
{\sl IRAS} observation.
At $850\,\mu$m, the upper limit is converted from the data reported by 
\citet{hunt05}.

The IR luminosity is slightly smaller than the previous prediction of T03.
We also find that the FIR peak wavelength is located at $\lambda 
\sim 60\;\mu$m, shorter than that of the T03 model SED.
In addition, the peak intensity is larger for the mixed model, since
the mixed model produce about 30~\% larger amount of dust than the unmixed 
model does (see Fig.~\ref{fig:dust_amount}). 
Since the system still remains optically thin at this age because of 
the large $r_{\rm SF}$, this difference is directly reflected to the difference
of the peak intensity.

The low opacity of \izw\ also leads to a low flux at N--MIR in the 
model SED in Figure~\ref{fig:sed_izw} both for unmixed and mixed cases
compared with the T03 model SED. 
As mentioned by T03, the NIR continuum of \izw\ is expected to
be dominated by stellar and nebular continuum radiation, and the contribution
of dust may be negligible.

We expect that these features will be found commonly in young 
galaxies if they are optically thin in the IR.
Hence, it may be useful to obtain some suggestions for the observational 
strategy of young galaxies at high-$z$.
We will discuss a well known real sample of such population of galaxies, 
Lyman-break galaxies, in the next subsection.

\subsection{Lyman-break galaxies}

Lyman-break galaxies (LBGs) are one of the most well-studied categories of
high-$z$ star-forming galaxies \citep[e.g.,][]{steidel99,steidel03}.
Even in LBGs, there is clear evidence that they contain 
non-negligible amount of dust \citep[e.g.,][]{adelberger00,calzetti01}. 
A high dust temperature ($\ga 70\,$K) is suggested by subsequent studies
\citep{ouchi99,chapman00,calzetti01,sawicki01}.

In order to make a consistent picture of the dust emission from LBGs, 
T04 applied the T03 model and made various predictions for these galaxies.
They also considered a power-law dust size distribution to examine its
effect on the resulting SEDs.
The T04 model reproduced the known observational properties of the 
dust emission from LBGs.

In this paper, we investigate the expected appearance of the LBGs with 
an improved dust grain formation of N03.

\subsubsection{Evolution of the SED of LBGs}\label{subsubsec:local}

\begin{figure*}
\centering\includegraphics[angle=90,width=0.9\textwidth]{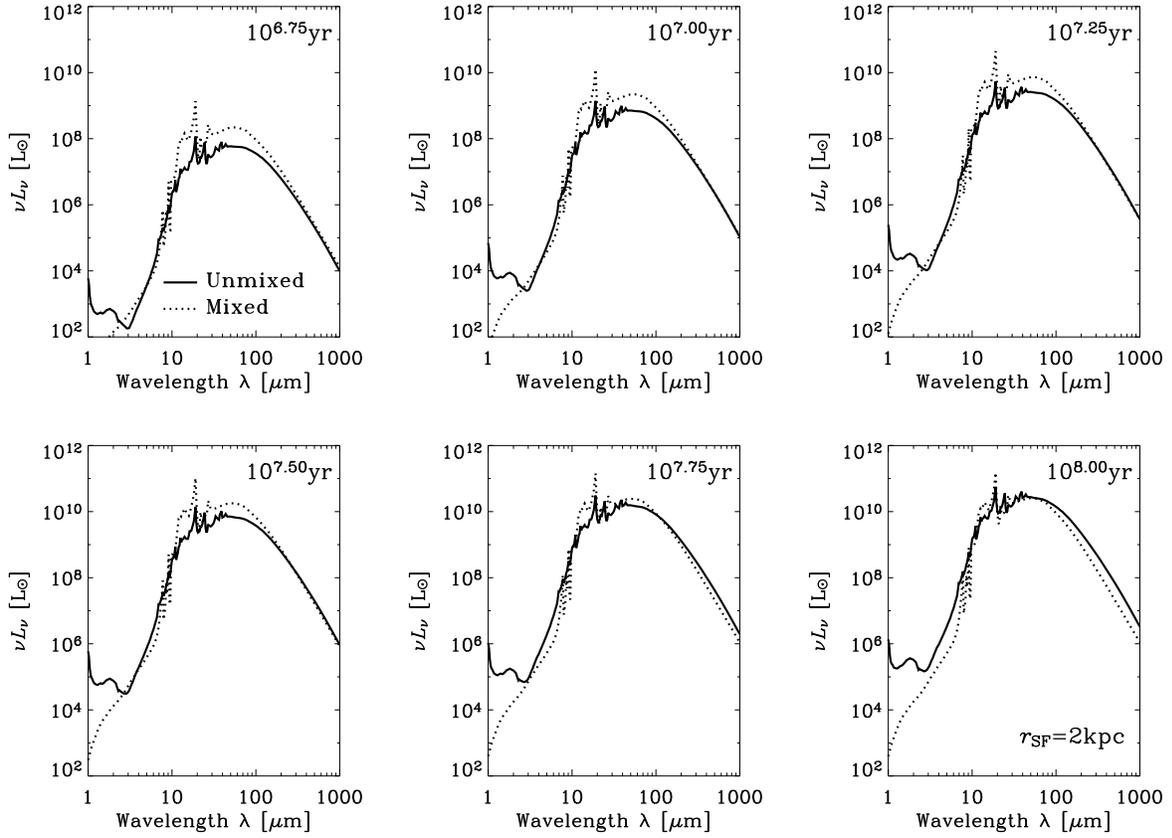}
\caption{The evolution of the SED of a typical Lyman-break galaxy (LBG).
The SFR is set to be $\mbox{SFR}=30\,M_\odot \mbox{yr}^{-1}$, and 
the size of the star-forming region $r_{\rm SF} = 2\,\mbox{kpc}$.
}\label{fig:lbg}
\end{figure*}

\begin{figure*}
\centering\includegraphics[angle=90,width=0.9\textwidth]{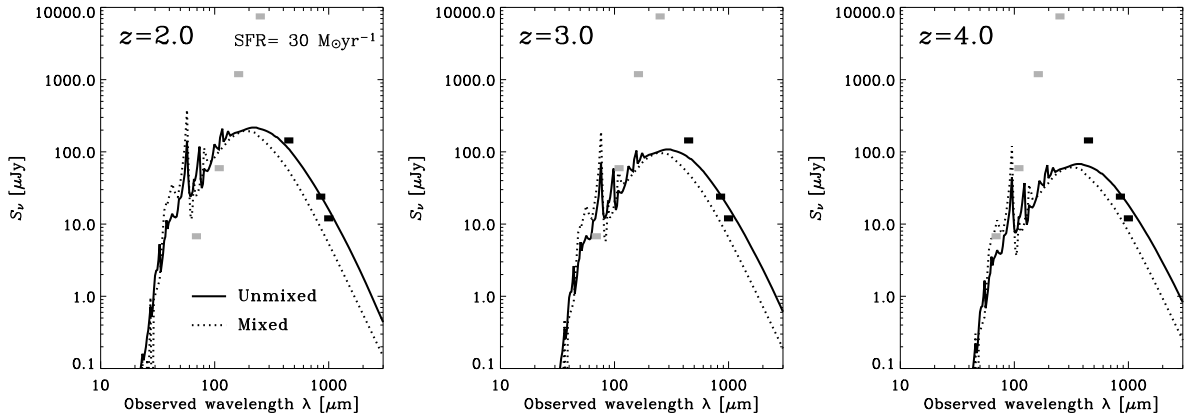}
\caption{The expected flux densities of an LBG.
In this figure we set the galaxy burst age of $t=10^8\,\mbox{yr}$.
Confusion limits of {\sl Herschel} and detection limits of ALMA 8-hour 
survey are also shown by gray and black thick horizontal lines, respectively.
}\label{fig:lbg_obs}
\end{figure*}

We set the input parameters of the SED model for LBGs as follows.
The SFR of LBGs spreads over the range of 
$\mbox{SFR} \simeq 1 \mbox{--} 300\,M_\odot \mbox{yr}^{-1}$ with a median 
of $\mbox{SFR} \simeq 20\,M_\odot \mbox{yr}^{-1}$ 
\citep[e.g., ][]{erb03}.
A constant SFR up to the age of $10^9$~yr is suggested to be
a good approximation \citep[e.g.,][]{baker04}.
Thus, the basic framework of the T03 model is also valid for LBGs.
In this work, we consider the moderate case of 
$\mbox{SFR}=30\,M_\odot \mbox{yr}^{-1}$ over the age of $10^{6.5}
\mbox{--}10^{8}$~yr.

The most important information to calculate the IR SED is the effective
size of the star forming region, but it is the most uncertain quantity
(see discussion of T04).
Since the mean half-light radius of LBGs is estimated to be $\sim 1.6\,
\mbox{kpc}$ from {\sl HST} observations \citep{erb03}, 
we use the galaxy radius as the radius of a star-forming region, 
and set $r_{\rm SF} = 2\,\mbox{kpc}$ according to T04.

The SED evolution of LBGs is shown in Figure~\ref{fig:lbg}.
Note that the scale in the ordinate of Figure~\ref{fig:lbg} 
is different from those in 
Figures~\ref{fig:sed_30pc} and \ref{fig:sed_100pc}.
Aside from luminosity, the behavior of the SED evolution of LBG is 
qualitatively very similar to that of a dwarf-like forming galaxy
with $r_{\rm SF}=100$~pc (i.e., a weak NIR continuum and a strong M--FIR 
emission is commonly seen in both).
This is because both the UV photon number density and dust number density 
are in the same order for an LBG and dwarf galaxy ($r_{\rm SF}=100$~pc).
Since the IR luminosity depends on the dust column density, the total
luminosity is different, but the shape of the SED is determined by the 
balance between the densities of UV photons and dust grains, they are
similar with each other as for the shape.

The difference between the SEDs of the unmixed and mixed cases is small, 
and they give almost the same result especially in the M--FIR.
This is explained as follows:
the most prominent difference between the two is the N--MIR continuum, 
which is produced mainly by stochastic heating of grains.
If the UV photon density is low, they are almost always in the lowest
temperature state because the incidence of a UV photon is less frequent.
In this case the N--MIR continuum from stochastically heated dust does not
contribute the continuum significantly, and the global SED shape is 
determined by larger grains in the equilibrium with the ambient UV radiation
field.
Under such condition, they are similar in shape with each other.
Thus, we see that it will be difficult to determine which 
scenario of dust production is correct from the observation of LBGs.

\subsubsection{Observability of the dust emission from LBG}

The flux density of a source at observed frequency, $\nu_{\rm obs}$
is obtained by 
\begin{eqnarray}\label{eq:flux}
  S_{\nu_{\rm obs}} =
    \frac{(1+z)L_{(1+z)\nu_{\rm obs}}}{4\pi d_{\rm L}(z)^2} 
    = \frac{(1+z)L_{\nu_{\rm em}}}{4\pi d_{\rm L}(z)^2} \,,
\end{eqnarray}
where $d_{\rm L}(z)$ is the luminosity distance corresponding to a 
redshift $z$, and $\nu_{\rm obs}$ and $\nu_{\rm em}$ are observed and emitted
frequency, respectively.

We show the observed IR/submm SEDs of LBGs at $z=2$, 3, and 4 in 
Figure~\ref{fig:lbg_obs}.
For simplicity we only show the SED with the age of $10^8$~yr.
The thick black short horizontal lines indicate the 3-$\sigma$ detection limits
for 8-hour observation by ALMA (Atacama Large Millimeter Array).\footnote{
URL: {\tt http://www.alma.info/}.}
Here we assumed 64 antennas and three wavelength bands, 450, 850, and 
$1080\,\mu$m.
We also show the 3-$\sigma$ source confusion limit of {\sl Herschel}\footnote{
URL: {\tt http://www.rssd.esa.int/herschel/}.
} 
at 75, 160, 250, and 350$\,\mu$m bands by thick gray horizontal lines.
These limits are based on `the photometric criterion' of \citep{lagache03}.
See also \citet[][]{ishii02,takeuchi04a}.

As discussed in T04, the detectability of LBGs is not strongly dependent on
their redshifts.
The predicted IR luminosity is factor of two smaller than that of T04, hence 
it is more difficult to detect LBGs than the discussion in T04:
we need roughly four times longer integration time than that derived from T04,
if we fix all the other conditions.
Detection at $350\;\mu$m seems impossible for moderate-SFR LBGs.
However at longer wavelengths, 
if the age $\ga 10^8\,{\rm yr}$ and $\mbox{SFR} \ga 30\,M_\odot\,
{\rm yr}^{-1}$, LBGs can be detected at a wide range of redshifts in the submm
by ALMA deep survey.
In the FIR, {\sl Herschel} will detect the dust emission from LBGs at 
$z\simeq 2$, but difficult at higher-$z$.
Since the SED does not differ for the unmixed or mixed dust formation,
our evaluation of the observability of LBGs
holds both for the mixed and unmixed dust formation.

\begin{figure*}
\centering\includegraphics[angle=90,width=0.9\textwidth]{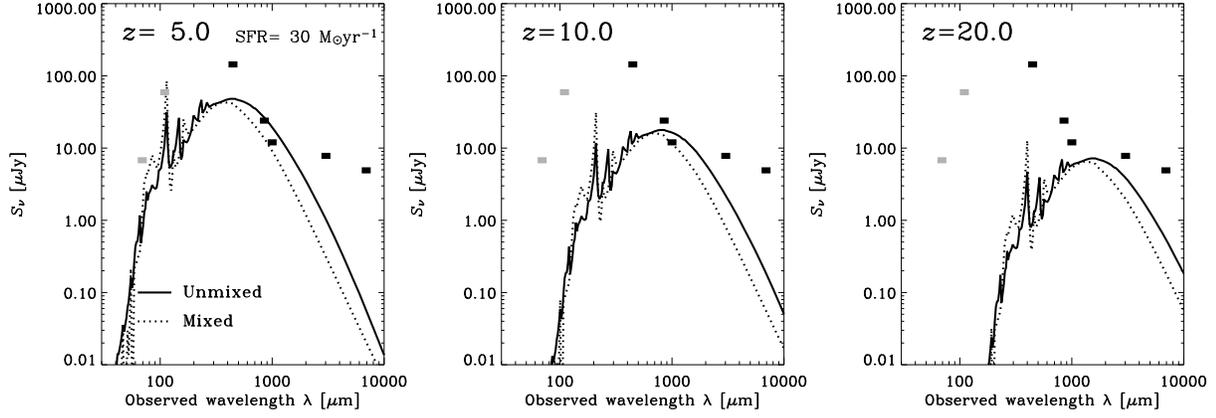}
\caption{The expected flux densities of an LBGs hypothetically located at
$z =5, 10$, and 20. 
The assumed parameters are the same as those for Figure~\ref{fig:lbg_obs}.
Solid lines represent the SEDs for the unmixed case of dust production, 
and dotted lines for the mixed case.
We put the same observational detection limits as Figure~\ref{fig:lbg_obs}
for {\sl Herschel} and ALMA.
}\label{fig:lbg_highz}
\end{figure*}

\subsection{Toward higher redshifts}

Based on the above discussions, we give a brief consideration on 
the observation of very high-$z$ galaxies ($z \ga 5$) here.
Direct observation of such galaxies are of vital importance to explore
the physics of galaxy formation.

First, assume a LBG at these redshifts.
Figure~\ref{fig:lbg_highz} shows such a situation.
We assumed the same physical parameters for LBGs.
{}From Figure~\ref{fig:lbg_highz}, we find that LBG-like objects can be 
detectable even at $z \simeq 10$ if $\mbox{SFR} \ga 30\; M_\odot
\mbox{yr}^{-1}$.
At $z \simeq 20$, it may be more difficult, but might be possible to
detect LBGs with very high SFR, by an ALMA ultra-deep survey with much 
longer integration time than eight hour.

However, it may not be reasonable to assume a galaxy like LBGs at $z \simeq 10$
because they are rather massive system in general 
($M_{\rm star} \simeq 10^{10}\,M_\odot$), and, from the modern cosmological 
viewpoint, such a massive system is very rare at such a high-$z$, young
universe.

\begin{figure*}
\centering\includegraphics[angle=90,width=0.9\textwidth]{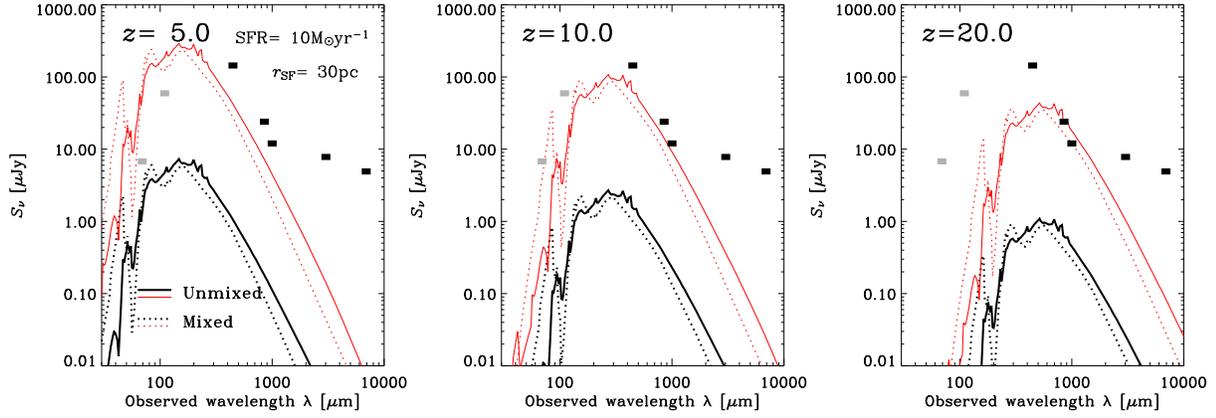}
\caption{The expected flux densities of a dwarf star-forming galaxy located at
$z =5, 10$, and 20. 
Solid lines represent the SEDs for the unmixed case of dust production, 
and dotted lines for the mixed case.
We assumed $\mbox{SFR}=10\,M_\odot\mbox{yr}^{-1}$, and 
$r_{\rm SF}=30\;\mbox{pc}$.
Thick lines are the expected SEDs for an unlensed galaxy, while the thin
lines are the ones for a gravitationally lensed galaxy by a magnification
factor of 40.
Again the detection limits are the same as those of Figure~\ref{fig:lbg_obs}.
}\label{fig:dwarf_highz_30pc}
\end{figure*}

\begin{figure*}
\centering\includegraphics[angle=90,width=0.9\textwidth]{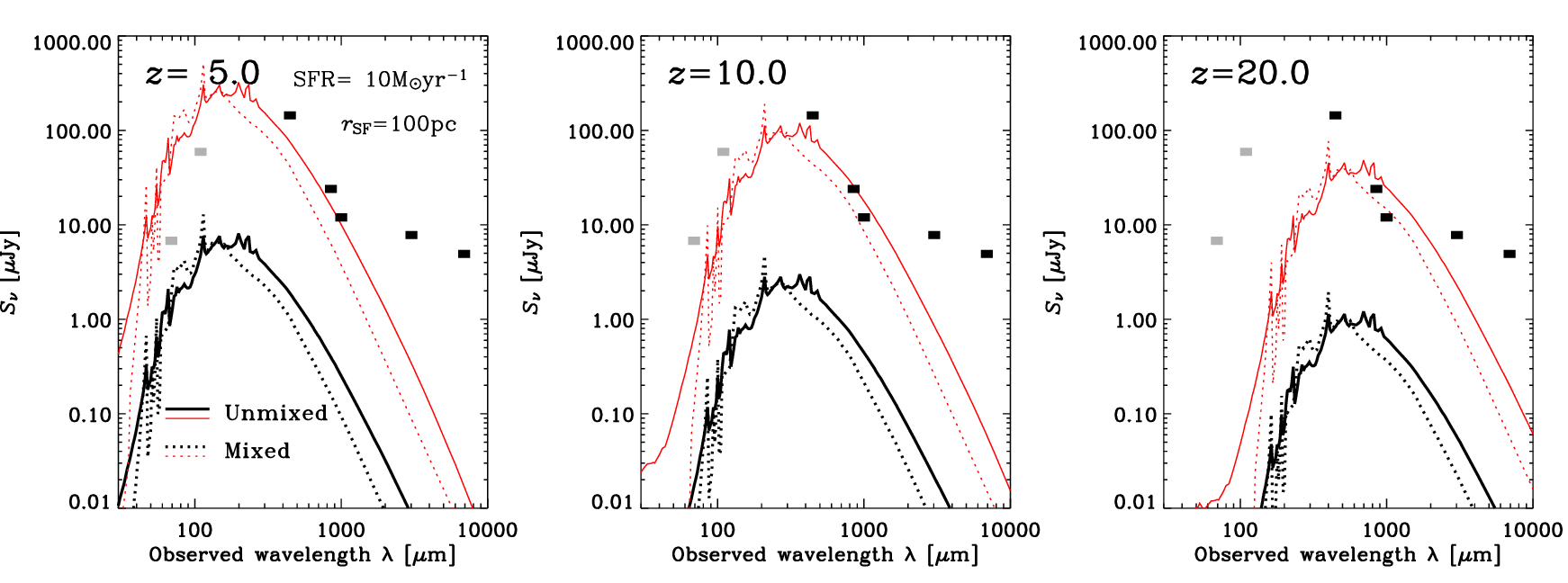}
\caption{The same as Figure~\ref{fig:dwarf_highz_30pc}, except that
$r_{\rm SF}=100\;\mbox{pc}$.
}\label{fig:dwarf_highz_100pc}
\end{figure*}

Here we discuss an object less massive than LBGs.
In modern hierarchical structure formation scenarios, it would be more 
reasonable to assume a small, subgalactic clump as a first forming galaxy.
Consider a dark halo of mass $\sim 10^9\;M_\odot$, then it is expected to 
contain a gas with mass $\simeq 10^8\;M_\odot$.
For this purpose, we calculate the SEDs for a dwarf star-forming galaxy.
If gas collapses on the free-fall timescale with an efficiency of 
$\epsilon_{\rm SF}$ (we assume $\epsilon_{\rm SF}=0.1$), we obtain
the following evaluation of the SFR \citet{hirashita04}:
\begin{eqnarray}
  \mbox{SFR} \simeq 0.1 \left(\frac{\epsilon_{\rm SF}}{0.1}\right)
    \left(\frac{M_{\rm gas}}{10^7\;M_\odot}\right)
    \left(\frac{n_{\rm H}}{100\;\mbox{cm}^{-3}}\right)^{1/2}\;
    [M_\odot \mbox{yr}^{-1}] \;,
\end{eqnarray}
and
\begin{eqnarray}
  n_{\rm H} \simeq 100 \left(\frac{r_{\rm SF}}{100\;\mbox{pc}}\right)^{-3}
    \left(\frac{M_{\rm gas}}{10^7\;M_\odot}\right) \;
    [\mbox{cm}^{-3}]\;,
\end{eqnarray}
where $n_{\rm H}$ is the hydrogen number density.
Then we have
\begin{eqnarray}
  \mbox{SFR} \simeq 0.1 \left(\frac{\epsilon_{\rm SF}}{0.1}\right)
    \left(\frac{M_{\rm gas}}{10^7\;M_\odot}\right)^{3/2}
    \left(\frac{r_{\rm SF}}{100\;\mbox{pc}}\right)^{-3/2} \;
    [M_\odot \mbox{yr}^{-1}]\;.
\end{eqnarray}
If we consider $M_{\rm gas}\simeq 10^8\;M_\odot$, we have $\mbox{SFR}\simeq
3(r_{\rm SF}/100\;\mbox{pc})^{-3/2}\;M_\odot\,\mbox{yr}^{-1}$.
In addition, as we see below, an extremely high-$z$ galaxy observed by 
{\sl HST} has a very compact morphology \citep{kneib04}.
We also mention that, from a theoretical side, high-$z$ galaxies are 
suggested to be dense and compact compared to nearby galaxies 
\citep{norman97,hirashita02b}.
Hence, it may be reasonable to assume a same
type of galaxy as local dwarfs discussed in Section~\ref{subsubsec:local}.
Thus, we consider a dwarf galaxy with $\mbox{SFR} = 10\;M_\odot\mbox{yr}^{-1}$
as an example, and we adopt $r_{\rm SF} = 30$~pc and 100~pc.
The age is set to be $10^7$~yr.
If the age is older, they will become easier to detect if a constant SFR takes
place.
We show the expected SEDs for such galaxies at $z=5$, 10, and 20 in 
Figures~\ref{fig:dwarf_highz_30pc} and \ref{fig:dwarf_highz_100pc}.
As expected, it seems almost impossible to detect such objects 
by {\sl Herschel} or ALMA.

There is, however, a hope to observe such a small forming galactic clump
directly: gravitational lensing works very well as a natural huge 
telescope.
For example, \cb\ is an ideal case of a lensed LBG \citep{nakanishi97,baker04},
and more recently, a small star-forming galaxy has been detected by
{\sl HST} and {\sl Spitzer} observations \citep[][]{kneib04,egami05}.
Hence, we can expect a forming galaxy which is intrinsically small and faint
but magnified by a gravitational potential of a cluster of galaxies.
If we assume a lens magnification factor of 40, such a small galaxy becomes
detectable.
This is depicted by the thin lines in Figures~\ref{fig:dwarf_highz_30pc}
and \ref{fig:dwarf_highz_100pc}.
Since the expected SED of such a compact dwarf galaxy has a strong MIR 
continuum at their rest frame, it can be feasible to detect at the FIR in 
the observed frame.
Indeed, at $z=5$ they appear above the confusion limits of {\sl Herschel}.
Since the confusion limit of {\sl Herschel} may be, in fact, very difficult to
reach in actual observations, a cooled FIR space telescope is more suitable 
for such observation, and this will be a strong scientific motivation for 
a future project like {\sl SPICA}.\footnote{URL: 
{\tt http://www.ir.isas.jaxa.jp/SPICA/index.html}.}
In the FIR wavelength regime ($50\mbox{--}200\;\mu\mbox{m}$) , the SEDs of 
unmixed and mixed cases might be distinguishable.
If a FIR spectroscopy is performed, which picture of dust production is 
plausible might be examined.
At higher-$z$, they can be detectable by ALMA survey.
Even at $z \simeq 20$, they can be detected by a standard 8-hour survey
of ALMA, if a lensing takes place.
It is interesting to note that, at $z\simeq 10$, the peak of their SED
locates at the wavelength range where the detection limits of both
{\sl Herschel} and ALMA are not very good.
A new effective (probably space) facility should be developed to overcome
this difficulty in observation.

For this line of study, we must estimate how frequently such lensing events 
occur for high-$z$ objects.
Suppose a cluster of galaxies at $z_{\rm l}\simeq 0.1\mbox{--}0.2$ 
whose dynamical mass $M_{\rm dyn}$ is $5 \times 10^{14}\,M_\odot$ 
and whose mass distribution obeys the singular isothermal sphere.
We denote the strong lensing cross section, i.e., the area of the region 
in the source plane for which the resulting magnification by a cluster 
is larger than $\mu$, as $\sigma(>\mu)$.
\citet[][]{perrotta02} presented $\sigma(>\mu)$ as a function of $M_{\rm dyn}$
for $z_{\rm lens}=1.0$.
Since $\sigma(>\mu)\propto {D_{\rm ls}}^2$ [$D_{\rm ls}$ is the 
angular-diameter distance between the lens and the source
\citep[e.g.,][]{asada98}] \citep{covone05}, 
we can convert their result to our condition and obtain 
$\sigma(>10) \simeq 30\;\mbox{arcsec}^2$ 
on the source plane.\footnote{On the image plane, 
the area corresponds to $\simeq \mu\sigma(>\mu)$.}
This result is almost independent of the source redshifts.
Setting the limiting flux density $S_\nu = 1\;\mu$Jy and using the number 
counts of \citet{hirashita02b} for galaxies at $z>5$, 
we have an expected number of galaxies 
suffering a strong lensing to be $\simeq 1\mbox{--}3$.
Thus, we expect at least a few strongly lensed IR galaxies to this survey
depth.

{}To have a concrete idea for the observational strategy, we must consider 
an important aspect of the lensing.
It is known that the gravitational lensing affects the galaxy number counts
in two opposite directions: i) it magnifies the flux density of galaxies, and
ii) it stretches the sky area and decreases their number density on the sky.
This is called the `magnification bias' \citep{fort97,bezecourt98}.
The unlensed number count $N_0(>S_\nu)$ is generally approximated by 
a power law 
\begin{eqnarray}
  N_0(>S_\nu) \propto {S_\nu}^\beta \;.
\end{eqnarray}
Then, by denoting the magnification factor as $\mu$, we obtain the lensed
number counts $N_{\rm lens} (>S_\nu)$ as
\begin{eqnarray}
  N_{\rm lens}(>S_\nu) = N_0(>S_\nu) \mu^{-(\beta+1)}\label{eq:lens}
\end{eqnarray}
\citep[e.g.,][]{fort97,takeuchi00}.
Equation~(\ref{eq:lens}) shows that whether the number counts increases
or decreases depends on the slope of the counts, $\beta$: if $\beta < -1$
lensing works as an enhancement of the counts.
Since the slope of the number counts of such galaxies is expected to be 
steeper than $-1$ \citep[e.g.,][]{hirashita02b}, the source density on 
the image plane will increase \citep[see also][]{perrotta02}.
Thus, we are confident that the lensing
will work as a really useful tool to detect forming small galaxies with
dust emission.

\section{Conclusion}\label{sec:conclusion}

Dust plays various important roles even in the very early phase of galaxy 
evolution \citep[e.g.,][]{hirashita02b}.
In such a young phase with a typical age less than $10^9$~yr, dust is 
predominantly supplied by SNe.
With the aid of a new physical model of dust production by SNe developed 
by N03, we constructed a model of dust emission from 
a very young galaxies according to T03.

N03 carefully took into account the radial density profile and the 
temperature evolution in the calculation of the dust formation in the ejecta 
of SNe II and PISNe.
They also showed that the produced dust species depends strongly on the mixing 
within SNe. 
We treated both unmixed and mixed cases and calculated the IR SED of young
galaxies for both cases.

The SEDs constructed from N03 dust production are less luminous than those
by T03 model by a factor of 2--3.
This difference is due to the improvement in the treatment of 
$Q(a,\lambda)$ at UV and the considered grain species.
The SED for the unmixed case is found to have a strong N--MIR continuum 
radiation in its early phase of the evolution ($\mbox{age} \la 10^{7.25}\;$yr) 
compared with that for the mixed case.
The N--MIR continuum is due to the emission from Si grains, which only exist in the 
species of the unmixed dust production.

We also calculated the IR extinction curves for young galaxies.
N03 dust gives a weaker extinction than that of T03 model because of the
small relative number of very small dust grains.
This is also the cause of the smaller IR luminosity of the present model.
For the unmixed case, NIR extinction is dominated by large grain of Si and 
amorphous carbon, 
and silicate features are less prominent compared to the curve given by T03.
On the contrary, the extinction curve of the mixed case has a similar shape
with that of T03.

The SED of a local starbursting dwarf galaxy, \sbs, was calculated.
Recent {\sl Spitzer} observations \citep{houck04} have implied a hotter dust
than previously thought. 
Our present model SED naturally reproduces the strong N--MIR continuum 
and the lack of cold FIR emission of \sbs.
We found that only the SED of unmixed case can reproduce the N--MIR continuum
of this galaxy.
Hence, as for \sbs, the unmixed dust production is preferred.
It will be interesting to proceed this line of study for higher-$z$ galaxies.

A prediction for the SED of another typical nearby star-forming 
dwarf galaxy, \izw, was then made.
Since \izw\ is dominated by very young star formation, we may adopt the 
model to this galaxy.
Using the dust species from N03 with this model, we find a weaker FIR 
emission than that of T03.
The N--MIR continuum is also expected to be much weaker than that of T03 SED.

We also calculated the evolution of the SED of LBGs.
For the parameters of LBGs, the unmixed and mixed picture does not affect
the appearance of the SED.
Hence, 
if we combine our model with the knowledge obtained from optical observations,
the observability of LBGs at submillimetre wavelengths are robust
conclusion independent of the details of the SNe dust production theory.

Finally, we considered the observations of forming galaxies at $z \ga 5$.
If there exist LBG-like galaxies at these redshifts, they can be detected
at $z \la 10$ by ALMA 8-hour survey 
if $\mbox{SFR} \ga 30\;M_\odot\mbox{yr}^{-1}$.
For small forming galaxies or subgalactic clumps, it is almost impossible to 
detect their intrinsic flux by ALMA or by {\sl Herschel}.
However, the gravitational lensing is found to be a very effective 
tool to detect such small star-forming galaxies at $z \ga 5$.
If we consider a compact dwarf galaxy with 
$\mbox{SFR}\simeq 10\;M_\odot\mbox{yr}^{-1}$ and $r_{\rm SF}\simeq 
30\;\mbox{pc}$, we can expect a few strongly magnified galaxies behind
a typical cluster of galaxies at $z \simeq 0.1\mbox{--}0.2$.

\section*{Acknowledgments}
First we are grateful to the anonymous referee for many insightful comments
which have improved the clarity of this paper much.
We deeply thank Akio K.\ Inoue, Jean-Paul Kneib and Giovanni Covone for 
illuminating discussions.
We made extensive use of the NASA Astrophysics Data System.
TTT, TTI, and HH have been supported by the Japan Society of the Promotion of 
Science.
TK is supported by a Grant--in--Aid for Scientific
Research from JSPS (16340051).

\appendix

\section{Details of the chemical evolution}\label{sec:chemical_evolution}

Here we present a detailed description of the chemical evolution we used 
in this work.
Full treatment of the present chemical evolution model including the feedback, 
dust destruction, and more complex SF history is given elsewhere
(Nozawa et al., in preparation).
Within the framework of the closed-box model, the total baryonic mass of 
a galaxy, $M_{\rm T}$, is conserved as
\begin{eqnarray}
  M_{\rm T} = M_{\rm gas}(0) = M_{\rm gas}(t) + M_{\rm dust}(t) 
    + M_{\rm star}(t) + M_{\rm rem}(t)\;,
\end{eqnarray}
where $M_{\rm gas}(t)$ is the gas mass, $M_{\rm dust}(t)$ is the dust mass,
$M_{\rm star}(t)$ is the stellar mass, and $M_{\rm rem}(t)$ is the mass of
stellar remnants in the galaxy at an age of $t$.
We start the calculation from homogeneous pristine gas, i.e., 
$M_{\rm dust}(0) = M_{\rm star}(0) = M_{\rm rem}(0) =0$.
The time evolution of the mass of interstellar medium, $M_{\rm ISM}(t) \equiv 
M_{\rm gas}(t) + M_{\rm dust}(t)$ is given by 
\begin{eqnarray}\label{eq:ev_ism}
  \left.\frac{dM_{\rm ISM}}{dt}\right|_t  = 
    -\psi(t) + \int_{m_{\rm l}}^{m_{\rm u}} \left[m-m_{\rm rem}(m)\right]
   \psi(t-\tau_m)\phi(m)dm \;,
\end{eqnarray}
where $\psi(t)$ is the SFR at an age $t$, $\phi(m)$ is the IMF normalized to 
unity in the mass interval 
$[m_{\rm l},\,m_{\rm u}]$ (subscripts l and u mean lower and upper mass
limit, respectively),
$m_{\rm rem}(m)$ is the mass of a single stellar remnant resulting from 
a progenitor star with mass $m$, and
$\tau_m$ is the lifetime of a star with mass $m$.

For the IMF, as we mentioned in the main text, we adopt the Salpeter IMF 
with mass range of $(m_{\rm l}, m_{\rm u})= (0.1\;M_\odot,100\;M_\odot)$.
The lifetime $\tau_m$ is evaluated by a fitting 
formula of zero-metallicity tracks with no mass loss, given by 
\citet{schaerer02}.
For SNe II (progenitor mass $8\mbox{--}40\;M_\odot$), we approximate 
the remnant mass $m_{\rm rem}$ by the linear fit based on 
the numerical result of SN explosions given by \citet{umeda02}:
\begin{eqnarray}\label{eq:remnant}
  m_{\rm rem}(m)= 0.06m + 0.93 \qquad(8\;M_\odot < m < 40\;M_\odot)\;.
\end{eqnarray}
For other progenitor mass ranges, we set $m_{\rm rem}(m)=0$ at $m < 8\;M_\odot$
(no remnants), and $m_{\rm rem}(m)=m$ at $40\;M_\odot < m < 100\;M_\odot$ 
(all the mass is swallowed by a black hole).
In this work, we do not treat PISNe in the calculation.
By Equation~(\ref{eq:remnant}), the time evolution of $M_{\rm rem}(t)$ is
expressed by 
\begin{eqnarray}\label{eq:ev_remnant}
  \left.\frac{dM_{\rm rem}}{dt}\right|_t = 
    \int_{m_{\rm l}}^{m_{\rm u}} m_{\rm rem}(m)\psi(t-\tau_m)\phi(m)dm \;.
\end{eqnarray}

The time evolution of $M_{\rm dust}(t)$ is, then, obtained as
\begin{eqnarray}\label{eq:ev_dust}
  \left.\frac{dM_{\rm dust}}{dt}\right|_t = 
    -\frac{M_{\rm dust}(t)}{M_{\rm ISM}(t)}\psi(t)+
    \int_{m_{\rm l}}^{m_{\rm u}} m_{\rm dust}(m)\psi(t-\tau_m)\phi(m)dm 
\end{eqnarray}
where $m_{\rm dust}(m)$ denotes the dust mass produced by a progenitor
with mass $m$.
We apply the dust formation model of N03 for $m_{\rm dust}$.
As we mentioned in \S\S\ref{subsec:dust_species}, we used the result of 
progenitor mass $20\;M_\odot$ for the fraction and the size distribution 
of dust grain species, hence the mass fraction and size distribution
are constant in time along with the evolution.

\section{Normalization of the number density of grain species}
\label{sec:normalization}

In this Appendix, we show how to obtain the normalization of
the dust size distribution $dN_i/da_i$,
\begin{eqnarray}\label{eq:dust_number_norm}
  \frac{dN_i}{da} da \equiv Af_i(a) da
\end{eqnarray}
where subscript $i$ denotes the species of dust, and $f_i$ 
is the mass fraction of dust of species $i$, given by N03, and
$A$ is the normalization constant.
We denote the material density of $i$-species dust by $\rho_i$,
and the SF region radius as $r_{\rm SF}$.
The total mass of dust species $i$, $M_i$, is then written as
\begin{eqnarray}
  M_i \hspace{-3mm}&=&\hspace{-3mm}
    \rho_i \int \frac{4 \pi a^3}{3} \frac{dN_i}{da} da
    = \rho_i \int \frac{4 \pi a^3}{3} A f_i(a) da \nonumber\\ 
  \hspace{-3mm}&=&\hspace{-3mm} \frac{4 \pi A}{3}\rho_i \int a^3 f_i(a) da \;.
\end{eqnarray}
The total dust mass of all the species, $M$, is 
\begin{eqnarray}
  M = \sum_i \frac{4 \pi A}{3} \rho_i \int f_i(a) a^3 da 
  = \frac{4 \pi A}{3}\sum_i \rho_i \int f_i(a) a^3 da \;.
\end{eqnarray}
Hence
\begin{eqnarray}
  A = \frac{3M}{\displaystyle 4 \pi \sum_i \rho_i \int f_i(a) a^3 da} \;.
\end{eqnarray}
Thus we can determine $A$ from the available data.
Note that only $M$ depends on time $t$, hence the time dependence of $A$ is 
given only through $M$.
The total number of grains, $N_i$, is expressed as 
\begin{eqnarray}\label{eq:dust_number}
  N_i = \int \frac{dN_i}{da_i}da_i \;.
\end{eqnarray}


\begin{thebibliography}{}
\bibitem[\protect\citeauthoryear{Adelberger \& Steidel}{2000}]{adelberger00}
 Adelberger K.\ L., Steidel C.\ C., 2000, ApJ, 544, 218

\bibitem[\protect\citeauthoryear{Arendt, Dwek, \& Moseley}{1999}]{arendt99}
 Arendt R.~G., Dwek E., Moseley S.~H., 1999, ApJ, 521, 234 

\bibitem[\protect\citeauthoryear{Arnett et al.}{1989}]{arnett89} 
 Arnett W.~D., Bahcall J.~N., Kirshner R.~P., Woosley S.~E., 1989, ARA\&A, 
 27, 629 

\bibitem[\protect\citeauthoryear{Asada}{1998}]{asada98}
 Asada H., 1998, ApJ, 501, 473 

\bibitem[\protect\citeauthoryear{Baker et al.}{2004}]{baker04}
 Baker A.\ J., Tacconi L.\ J., Genzel R., Lehnert M.\ D., Lutz D.,
 2004, ApJ, 604, 125

\bibitem[\protect\citeauthoryear{Begemann et al.}{1991}]{begemann91}
 Begemann B., Dorschner J., Henning T., Mutschke H., Thamm, E., 1994, 
 ApJ. 423, L71

\bibitem[\protect\citeauthoryear{Bertoldi et al.}{2003}]{bertoldi03}
 Bertoldi F., Carilli C.\ L., Cox P., Fan X., Strauss M.\ A., Beelen A., 
 Omont A., Zylka R., 2003, A\&A, 406, L55

\bibitem[\protect\citeauthoryear{B\'{e}zecourt, Pell\'{o}, \& 
Soucail}{1998}]{bezecourt98}
 B\'{e}zecourt J., Pell\'{o} R., Soucail G., 1998, A\&A, 330, 399 

\bibitem[\protect\citeauthoryear{Bohren \& Huffman}{1983}]{bohren83}
 Bohren C.\ F., Huffman D.\ R., 1983, Absorption and Scattering of Light by
 Small Particles, Wiley, New York

\bibitem[\protect\citeauthoryear{Bouwens et al.}{2004a}]{bouwens04a}
 Bouwens R.~J., et al., 2004a, ApJ, 606, L25 

\bibitem[\protect\citeauthoryear{Bouwens et al.}{2004b}]{bouwens04b}
 Bouwens R.~J., et al., 2004b, ApJ, 616, L79 

\bibitem[\protect\citeauthoryear{Calzetti}{2001}]{calzetti01}
 Calzetti D., 2001, PASP, 113, 1449

\bibitem[\protect\citeauthoryear{Calzetti et al.}{2000}]{calzetti00}
 Calzetti  D., Armus  L., Bohlin  R.\ C., Kinney  A.\ L., Koornneef  J., 
 Storchi-Bergmann T., 2000, ApJ, 533, 682

\bibitem[\protect\citeauthoryear{Cannon et al.}{2002}]{cannon02}
 Cannon J.\ M., Skillman E.\ D., Garnett D.\ R., Dufour R.\ J.,
 2002, ApJ, 565, 931

\bibitem[\protect\citeauthoryear{Cardelli, Clayton, \& Mathis}
{Cardelli et al.}{1989}]{cardelli89}
 Cardelli J.\ A., Clayton G.\ C., Mathis J.\ S., 1989, ApJ, 345, 245

\bibitem[\protect\citeauthoryear{Chapman et al.}{2000}]{chapman00}
 Chapman S.\ C., et al., 2000, MNRAS, 319, 318

\bibitem[\protect\citeauthoryear{Covone, Sereno, \& de Ritis}{2005}]{covone05}
 Covone G., Sereno M., de Ritis R., 2005, MNRAS, 357, 773

\bibitem[\protect\citeauthoryear{Dale et al.}{2001}]{dale01} 
 Dale D.\ A., Helou G., Neugebauer G., Soifer B.\ T., Frayer D.\ T., 
 Condon J.\ J., 2001b, AJ, 122, 1736

\bibitem[\protect\citeauthoryear{Dorschner et al.}{1995}]{dorschner95}
 Dorschner J., Begemann B., Henning Th., Jaeger C., Mutschke H., 1995, 
 A\&A, 300, 503

\bibitem[\protect\citeauthoryear{Douvion et al.}{2001}]{douvion01}
 Douvion T., Lagage P.~O., Cesarsky C.~J., Dwek E., 2001, A\&A, 373, 281 

\bibitem[\protect\citeauthoryear{Draine \& Lee}{1984}]{draine84}
 Draine B.\ T., Lee H.\ M., 1984, ApJ, 285, 89

\bibitem[\protect\citeauthoryear{Draine \& Anderson}{1985}]{draine85}
 Draine B.\ T., Anderson L., 1985, ApJ, 292, 494

\bibitem[\protect\citeauthoryear{Draine \& Li}{2001}]{draine01}
 Draine B.\ T., Li A., 2001, ApJ, 551, 807

\bibitem[\protect\citeauthoryear{Dunne et al.}{2003}]{dunne03} 
 Dunne L., Eales S., Ivison R., Morgan H., Edmunds M., 2003, Natur, 424, 285 

\bibitem[\protect\citeauthoryear{Dwek \& Scalo}{1980}]{dwek80}
 Dwek E., Scalo J.\ M., 1980, ApJ, 239, 193

\bibitem[\protect\citeauthoryear{Dwek}{1998}]{dwek98}
 Dwek E., 1998, ApJ, 501, 643

\bibitem[\protect\citeauthoryear{Dwek}{2004}]{dwek04}
 Dwek E., 2004, ApJ, 607, 848 

\bibitem[\protect\citeauthoryear{Eales et al.}{2003}]{eales03}
 Eales S., Bertoldi F., Ivison R., Carilli C., Dunne L., Owen F.,
 2003, MNRAS, 344, 169

\bibitem[\protect\citeauthoryear{Edo}{1985}]{edo85}
 Edo O., 1983, PhD Dissertation, Dept.\ of Physics, University of Arisona

\bibitem[\protect\citeauthoryear{Edward}{1985}]{edward85}
 Edward D.\ F., 1985, in Handbook of Optical Constants of Solids
 ed.\ E.\ D.\ Palik, Academic Press, San Diego, USA p.547

\bibitem[\protect\citeauthoryear{Egami et al.}{2005}]{egami05} 
 Egami E., et al., 2005, ApJ, 618, L5 

\bibitem[\protect\citeauthoryear{Erb et al.}{2003}]{erb03}
 Erb D.\ K., et al., 2003, ApJ, 591, 101

\bibitem[\protect\citeauthoryear{Feder et al.}{1966}]{feder66}
 Feder J., Russell K.\ C., Lothe J., Pound M., 1966, Adv.\ in Phys., 15, 111

\bibitem[\protect\citeauthoryear{Fort, Mellier, \& Dantel-Fort}{1997}]{fort97}
 Fort B., Mellier Y., Dantel-Fort M., 1997, A\&A, 321, 353 

\bibitem[\protect\citeauthoryear{Gail, Keller, \& Sedlmayr}
{Gail et al.}{1984}]{gail84}
 Gail H.-P., Keller R., Sedlmayr E., 1984, A\&A, 133, 320

\bibitem[\protect\citeauthoryear{Gail \& Sedlmayr}{1979}]{gail79}
 Gail H.-P., Sedlmayr E., 1979, A\&A, 77, 165

\bibitem[\protect\citeauthoryear{Galliano et al.}{2003}]{galliano03}
 Galliano F., Madden S.~C., Jones A.~P., Wilson C.~D., Bernard J.-P.,
 Le Peintre F., 2003, A\&A, 407, 159 

\bibitem[\protect\citeauthoryear{Galliano et al.}{2005}]{galliano05}
 Galliano F., Madden S.~C., Jones A.~P., Wilson C.~D., Bernard J.-P.,
 2005, A\&A, 434, 867

\bibitem[\protect\citeauthoryear{Green, Tuffs, \& Popescu}
{Green et al.}{2004}]{green04}
 Green D.~A., Tuffs R.~J., Popescu C.~C., 2004, MNRAS, 355, 1315 

\bibitem[\protect\citeauthoryear{Gr{\o}nvold et al.}{1991}]{gronvold91}
 Gr{\o}nvold F., St{\o}len S., Labban A. K., Westrum E. F., Jr., 1991, 
 J.\ Chem. Thermodynamics, 23, 261

\bibitem[\protect\citeauthoryear{Hama \& Suito}{1999}]{hama99}
 Hama J., Suito K., 1999, Physics of the Earth and Planetary Interiors,
 114, 165

\bibitem[\protect\citeauthoryear{Heger \& Woosley}{2002}]{heger02}
 Heger A., Woosley S.\ E., 2002, ApJ, 567, 532

\bibitem[\protect\citeauthoryear{Hines et al.}{2004}]{hines04} 
 Hines D.~C., et al., 2004, ApJS, 154, 290 

\bibitem[\protect\citeauthoryear{Hirashita, Hunt, \& Ferrara}
{Hirashita et al.}{2002}]{hirashita02a} 
 Hirashita H., Hunt L.\ K., Ferrara A., 2002, MNRAS, 330, L19 

\bibitem[\protect\citeauthoryear{Hirashita \& Ferrara}{2002}]{hirashita02b} 
 Hirashita H., Ferrara A., 2002, MNRAS, 337, 921

\bibitem[\protect\citeauthoryear{Hirashita \& Hunt}{2004}]{hirashita04} 
 Hirashita H., Hunt L.\ K., 2004, A\&A, 421, 555

\bibitem[\protect\citeauthoryear{Hirashita et al.}{2005}]{hirashita05} 
 Hirashita H., Nozawa T., Kozasa T., Ishii T.\ T., Takeuchi T.\ T., 2005, 
 MNRAS, 357, 1077

\bibitem[\protect\citeauthoryear{Houck et al.}{2004}]{houck04}
 Houck J.\ R.\ et al., 2004, ApJS, 154, 211

\bibitem[\protect\citeauthoryear{Hughes et al.}{1998}]{hughes98}
 Hughes D.\ H., et al., 1998, Natur, 394, 241

\bibitem[\protect\citeauthoryear{Hunt, Vanzi, \& Thuan}
{Hunt et al.}{2001}]{hunt01}
 Hunt L.\ K., Vanzi L., Thuan T.\ X., 2001, ApJ, 377, 66

\bibitem[\protect\citeauthoryear{Hunt, Giovanardi, \& Helou}
{Hunt et al.}{2002}]{hunt02}
 Hunt L.\ K., Giovanardi C., Helou G., 2002, A\&A, 394, 873

\bibitem[\protect\citeauthoryear{Hunt, Thuan, \& Izotov}
{Hunt et al.}{2003}]{hunt03}
 Hunt L.\ K., Thuan T.\ X., Izotov Y.\ I., 2003, ApJ, 2003, 588, 281

\bibitem[\protect\citeauthoryear{Hunt et al.}{2004}]{hunt04} 
 Hunt L.~K., Dyer K.~K., Thuan T.~X., Ulvestad J.~S., 2004, ApJ, 606, 853 

\bibitem[\protect\citeauthoryear{Hunt, Bianchi, \& Maiolino}
{Hunt et al.}{2005}]{hunt05}
 Hunt L.\ K., Bianchi S., \& Maiolino R., 2005, A\&A, 434, 849

\bibitem[\protect\citeauthoryear{Inoue, Hirashita, \& Kamaya}{2000}]{inoue00}
 Inoue A.~K., Hirashita H., Kamaya H., 2000, PASJ, 52, 539 

\bibitem[\protect\citeauthoryear{Inoue}{2005}]{inoue05}
 Inoue, A.\ K., 2005, MNRAS, 359, 171

\bibitem[\protect\citeauthoryear{Ishii, Takeuchi, \& Sohn}
{Ishii et al.}{2002}]{ishii02}
 Ishii T.\ T., Takeuchi T.\ T., Sohn J.-J., 2002, in Infrared and
 Submillimeter Space Astronomy: An International Colloquium to Honor the
 Memory of Guy Serra, p.169

\bibitem[\protect\citeauthoryear{J\"{a}ger et al.}{1995}]{jager03}
 J\"{a}ger C., Dorschner J., Mutschke H., Posch Th., Henning Th., 2003, 
 A\&A, 408, 193

\bibitem[\protect\citeauthoryear{Jones, Tielens, \& Hollenbach}
{Jones et al.}{1996}]{jones96}
 Jones A.\ P., Tielens A.\ G.\ G.\ M., Hollenbach D.\ J., 1996, ApJ, 469, 740

\bibitem[\protect\citeauthoryear{Joy \& Lester}{1988}]{joy88}
 Joy M., Lester D.\ F., 1988, ApJ, 331, 1451

\bibitem[\protect\citeauthoryear{Kneib et al.}{2004}]{kneib04} 
 Kneib J., Ellis R.~S., Santos M.~R., Richard J., 2004, ApJ, 607, 697 

\bibitem[\protect\citeauthoryear{Krause et al.}{2004}]{krause04} 
 Krause O., Birkmann S.~M., Rieke G.~H., Lemke D., Klaas U., Hines D.~C., 
 Gordon K.~D., 2004, Natur, 432, 596 

\bibitem[\protect\citeauthoryear{Kozasa \& Hasegawa}{1987}]{kozasa87}
 Kozasa T., Hasegawa H., 1987, Prog.\ Theor.\ Phys., 77, 1402

\bibitem[\protect\citeauthoryear{Kozasa, Hasegawa, \& Nomoto}
{Kozasa et al.}{1989}]{kozasa89}
 Kozasa T., Hasegawa H., Nomoto K., 1989, ApJ, 344, 325

\bibitem[\protect\citeauthoryear{Kr\"ugel}{2003}]{krugel03}
 Kr\"ugel E., 2003, The Physics of Interstellar Dust, Institute of Physics
 Publishing, Bristol

\bibitem[\protect\citeauthoryear{Lagache, Dole, \& Puget}
{Lagache et al.}{2003}]{lagache03}
 Lagache, G., Dole, H., Puget, J.-L., 2003, MNRAS, 338, 555

\bibitem[\protect\citeauthoryear{Ledoux, Bergeron, \& Petitjean}
{Ledoux et al.}{2002}]{ledoux02}
 Ledoux C., Bergeron J., Petitjean P., 2002, A\&A, 385, 802

\bibitem[\protect\citeauthoryear{Ledoux, Petitjean, \& Srianand}
{Ledoux et al.}{2003}]{ledoux03}
 Ledoux C., Petitjean P., Srianand R., 2003, MNRAS, 346, 209

\bibitem[\protect\citeauthoryear{Lilly et al.}{1996}]{lilly96}
 Lilly S.\ J., Le F\`{e}vre O., Hammer F., Crampton D., 1996, ApJ, 460, L1

\bibitem[\protect\citeauthoryear{Lynch \& Hunter}{1991}]{lynch91}
 Lynch, D.\ W., Hunter W.\ R., 1991, in Handbook of Optical Constants of
 Solids II, ed.\ E. D. Palik, Academic Press, San Diego, p.388

\bibitem[\protect\citeauthoryear{Madau et al.}{1996}]{madau96}
 Madau P., Ferguson H.\ C., Dickinson M.\ E., Giavalisco M., Steidel C.\ C., 
 Fruchter A., 1996, MNRAS, 283, 1388

\bibitem[\protect\citeauthoryear{Maiolino et al.}{2004}]{maiolino04}
 Maiolino R., Schneider R., Oliva E., Bianchi S., Ferrara A., Mannucci F., 
 Pedani M., Roca Sogorb M., 2004, Natur, 431, 533 

\bibitem[\protect\citeauthoryear{Mathis, Rumpl, \& Nordsieck}{1977}]{mathis77}
 Mathis J.\ S., Rumpl W., Nordsieck K.\ H., 1977, ApJ, 217, 425

\bibitem[\protect\citeauthoryear{Mezzasalma}{2000}]{mezzasalma00}
 Mezzasalma S. A., 2000, Journal of Physics and Chemistry of Solids, 61, 593

\bibitem[\protect\citeauthoryear{Morgan \& Edmunds}{2003}]{morgan03b}
 Morgan H.~L., Edmunds M.~G., 2003, MNRAS, 343, 427 

\bibitem[\protect\citeauthoryear{Morgan et al.}{2003}]{morgan03a} 
 Morgan H.~L., Dunne L., Eales S.~A., Ivison R.~J., Edmunds M.~G., 2003, 
 ApJ, 597, L33 

\bibitem[\protect\citeauthoryear{Mukai}{1989}]{mukai89}
 Mukai T., 1989, in Evolution of Interstellar Dust and Related Topics, 
 eds., A.\ Bonetti, J.\ M.\ Greenberg, S.\ Aiello,  Elsevier
 Science Publishers, Amsterdam, p.397

\bibitem[\protect\citeauthoryear{Nakanishi et al.}{1997}]{nakanishi97}
 Nakanishi K., Ohta K., Takeuchi T.~T., Akiyama M., Yamada T., Shioya Y., 
 1997, PASJ, 49, 535 

\bibitem[\protect\citeauthoryear{Norman \& Spaans}{1997}]{norman97}
 Norman C.~A., Spaans M., 1997, ApJ, 480, 145 

\bibitem[\protect\citeauthoryear{Nozawa et al.}{2003}]{nozawa03}
 Nozawa T., Kozasa T., Umeda H., Maeda K., Nomoto K., 2003, ApJ, 598, 785 (N03)

\bibitem[\protect\citeauthoryear{Oganov, Brodhold, \& Price}{2000}]{oganov00}
 Oganov A. R., Brodholt J. P., Price G. D., 2000, 
 Physics of the Earth and Planetary Interiors, 122, 277

\bibitem[\protect\citeauthoryear{\"{O}stlin}{2000}]{ostlin00}
 \"{O}stlin G., 2000, ApJ, 535, L99

\bibitem[\protect\citeauthoryear{Ouchi et al.}{1999}]{ouchi99}
 Ouchi M., Yamada T., Kawai H., Ohta K., 1999, ApJ, 517, L19

\bibitem[\protect\citeauthoryear{Perrotta et al.}{2002}]{perrotta02}
 Perrotta F., Baccigalupi C., Bartelmann M., De Zotti G., Granato G.~L.,
 2002, MNRAS, 329, 445 

\bibitem[\protect\citeauthoryear{Philipp}{1985}]{philipp85}
 Philipp H. R., 1985, in Handbook of Optical Constants of Solids, 
 ed.\ E.\ D.\ Palik, Academic Press, San Diego, p.719

\bibitem[\protect\citeauthoryear{Plante \& Sauvage}{2002}]{plante02}
 Plante S., Sauvage M., 2002, AJ, 124, 1995

\bibitem[\protect\citeauthoryear{Roessler \& Huffman}{1991}]{roessler91}
 Roessler D.\ M., Huffman D.\ R., 1991, in Handbook of Optical Constants of
 Solids II, ed.\ E. D. Palik, Academic Press, San Diego, CA, USA p.919

\bibitem[\protect\citeauthoryear{Salpeter}{1955}]{salpeter55}
 Salpeter E., 1955, ApJ, 121, 161

\bibitem[\protect\citeauthoryear{Sawicki}{2001}]{sawicki01}
 Sawicki M., 2001, AJ, 121, 2405

\bibitem[\protect\citeauthoryear{Schaerer}{2002}]{schaerer02}
 Schaerer D., 2002, A\&A, 382, 28

\bibitem[\protect\citeauthoryear{Schneider, Ferrara \& Salvaterra}{2004}]{schneider04}
 Schneider R., Ferrara A., Salvaterra R., 2004, MNRAS, 351, 1379

\bibitem[\protect\citeauthoryear{Semenov et al.}{2003}]{semenov03}
 Semenov D., Henning Th., Helling Ch., Ilgner M., Sedlmayr E., 2003, 
 A\&A, 410, 611

\bibitem[\protect\citeauthoryear{Shepherd et al.}{1991}]{shepherd91}
 Shepherd J.~P., Koenitzer J.~W., Arag{\'o}n R., Spa{\l}ek J., Honig J.~M., 
 1991, PhRvB, 43, 8461 

\bibitem[\protect\citeauthoryear{Siethoff \& Ahlborn}{1996}]{siethoff96}
 Siethoff H., Ahlborn K., 1996, J.\ Appl. Phys., 79, 2968

\bibitem[\protect\citeauthoryear{Stanway, Bunker, \& McMahon}
{Stanway et al.}{2003}]{stanway03}
 Stanway E.\ R., Bunker A.\ J., McMahon G., 2003, MNRAS, 342, 439

\bibitem[\protect\citeauthoryear{Steidel et al.}{1996}]{steidel96}
 Steidel C.\ C., Giavalisco M., Pettini M., Dickinson M., Adelberger K.\ L.,
 1996, ApJ, 462, L17

\bibitem[\protect\citeauthoryear{Steidel et al.}{1999}]{steidel99}
 Steidel C.\ C., Adelberger K.\ L., Giavalisco M., Dickinson M., 
 Pettini M., 1999, ApJ, 519, 1

\bibitem[\protect\citeauthoryear{Steidel et al.}{2003}]{steidel03} 
 Steidel C.\ C., et al., 2003, ApJ, 592, 728


\bibitem[\protect\citeauthoryear{Takeuchi, Yoshikawa, \& Yonehara}
{2000}]{takeuchi00}
 Takeuchi T.\ T., Yoshikawa K., Yonehara, A., 2000, in Mid- and Far-Infrared 
 Astronomy and Space Missions, ed.\ T.\ Matsumoto \& H.\ Shibai,
 ISAS Report, SP-14, 163

\bibitem[\protect\citeauthoryear{Takeuchi et al.}{2001a}]{takeuchi01a}
 Takeuchi T.\ T., Ishii T.\ T., Hirashita H., Yoshikawa K., Matsuhara H., 
 Kawara K., Okuda H., 2001a, PASJ, 53, 37

\bibitem[\protect\citeauthoryear{Takeuchi et al.}{2001b}]{takeuchi01b}
 Takeuchi T.\ T., Kawabe R., Kohno K., Nakanishi K., Ishii T.\ T., 
 Hirashita H., Yoshikawa K., 2001b, PASP, 113.\ 586

\bibitem[\protect\citeauthoryear{Takeuchi et al.}{2003}]{takeuchi03a}
 Takeuchi T.\ T., Hirashita H., Ishii T.\ T., Hunt L.\ K., 
 Ferrara A., 2003, MNRAS, 343, 839 (T03)

\bibitem[\protect\citeauthoryear{Takeuchi, Yoshikawa, \& Ishii}{2003}]{takeuchi03b}
 Takeuchi, T.\ T., Yoshikawa, K., Ishii, T.\ T.\ 2003, ApJ, 587, L89

\bibitem[\protect\citeauthoryear{Takeuchi \& Ishii}{2004a}]{takeuchi04a}
 Takeuchi T.\ T., Ishii T.\ T., 2004a, ApJ, 604, 40

\bibitem[\protect\citeauthoryear{Takeuchi \& Ishii}{2004}]{takeuchi04b}
 Takeuchi T.\ T., Ishii T.\ T., 2004b, A\&A, 426, 425 (T04)

\bibitem[\protect\citeauthoryear{Takeuchi et al.}{2005}]{takeuchi05}
 Takeuchi T.\ T., Buat V., Iglesias-P\'{a}ramo J., Boselli A., 
 Burgarella D., 2005, A\&A, 432, 423

\bibitem[\protect\citeauthoryear{Tanaka, Tanaka, \& Nakazawa}
{Tanaka et al.}{2002}]{tanaka02}
 Tanaka K.\ K., Tanaka H., Nakazawa K., 2002, Icarus, 160, 197

\bibitem[\protect\citeauthoryear{Thuan, Sauvage, \& Madden}{1999}]{thuan99}
 Thuan T.\ X., Sauvage M., Madden S., 1999, ApJ, 516, 783

\bibitem[\protect\citeauthoryear{Tinsley \& Danly}{1980}]{tinsley80}
 Tinsley B.\ M., Danly, L., 1980, ApJ, 242, 435

\bibitem[\protect\citeauthoryear{Todini \& Ferrara}{2001}]{todini01}
 Todini P., Ferrara A., 2001, MNRAS, 325, 726 (TF01)

\bibitem[\protect\citeauthoryear{Toon, Pollack, \& Khare}
{Toon et al.}{1976}]{toon76}
 Toon O.\ B., Pollack J.\ B., Khare B.\ N., 1976, J. Geophys.\ Res., 81, 5733

\bibitem[\protect\citeauthoryear{Totani \& Takeuchi}{2002}]{totani02}
 Totani T., Takeuchi T.\ T., 2002, ApJ, 570, 470

\bibitem[\protect\citeauthoryear{Umeda \& Nomoto}{2002}]{umeda02}
 Umeda H., Nomoto, K., 2002, ApJ, 565,385

\bibitem[\protect\citeauthoryear{Vanzi et al.}{2000}]{vanzi00}
 Vanzi L., Hunt L.\ K., Thuan T.\ X., Izotov Y.\ I., 2000, A\&A, 363, 493

\bibitem[\protect\citeauthoryear{Whittet}{1992}]{whittet92}
 Whittet D.\ C.\ B., 1992, Dust in the Galactic Environment, IOP, New York

\bibitem[\protect\citeauthoryear{Wilson \& Batrla}{2005}]{wilson05}
 Wilson T.\ L., Batrla W., 2005, A\&A, 430, 561
\end{thebibliography}
\end{document}